\documentstyle{article}

\font\tenrm=cmr10
\font\tenit=cmti10
\font\elevenbf=cmbx10 scaled\magstep 1
\font\elevenrm=cmr10 scaled\magstep 1
\font\elevenit=cmti10 scaled\magstep 1

\def\l{\stackrel{{\textstyle<}}{\sim}}
\def\r{\stackrel{{\textstyle>}}{\sim}}

\renewenvironment{thebibliography}[1]
 { \elevenrm
   \begin{list}{\arabic{enumi}.}
    {\usecounter{enumi} \setlength{\parsep}{0pt}
     \setlength{\itemsep}{3pt} \settowidth{\labelwidth}{#1.}
     \sloppy
    }}{\end{list}}

\begin{document}
\begin{flushright}
CTP--TAMU--52/93 \\
NUB--TH--3073-93 \\
SSCL--Preprint--503
\end{flushright}
\begin{center}
\vglue 10pt
{\elevenbf SUPERSYMMETRY AND SUPERGRAVITY:}
\vglue 3pt
{\elevenbf PHENOMENOLOGY AND GRAND UNIFICATION}
\vglue 5pt
{\tenrm R. ARNOWITT\footnote{Lectures at VII J. A. Swieca Summer School, Campos
do Jordao,
Brazil, 1993}$^{ab}$ and PRAN NATH$^c$\\}
\baselineskip=13pt
{\tenit \footnote{Permanent Address}$^a$Center for Theoretical Physics,
Department of Physics\\}
{\tenit Texas A\&M University, College Station, TX  77843-4242\\}
{\tenit $^b$Physics Research Division, Superconducting Super\\}
{\tenit Collider Laboratory, Dallas, TX  75237\\}
\baselineskip=12pt
{\tenit $^c$Department of Physics, Northeastern University\\}
{\tenit Boston, MA  02115\\}
\vglue 0.8cm
{\tenrm ABSTRACT}
\end{center}
\vglue 0.3cm
{\rightskip=3pc
\leftskip=3pc
\tenrm\baselineskip=12pt
\noindent
}
A survey is given of supersymmetry and supergravity and their phenomenology.
Some of
the topics discussed are the basic ideas of global supersymmetry, the minimal
supersymmetric Standard Model (MSSM) and its phenomenology, the basic ideas of
local supersymmetry (supergravity), grand unification, supersymmetry breaking
in
supergravity grand unified models, radiative breaking of $SU(2) \times U(1)$,
proton decay, cosmological constraints, and predictions of supergravity grand
unified models.  While the number of detailed derivations are necessarily
limited, a sufficient number of results are given so that a reader can get a
working knowledge of this field.

\vglue 0.6cm
\elevenrm
\baselineskip=14pt
{\elevenbf\noindent 1. Introduction}
\vglue 0.2cm

The Standard Model (SM) with three generations of quarks and leptons, based on
the gauge group $SU(3) \times SU(2) \times U(1)$ and possessing one Higgs
doublet to realize electroweak breaking, is in excellent agreement with all
current data.  In fact, the Standard Model is one of the most successful
theories ever constructed, and at present accounts for all microscopic
physics.  In spite of this, there is a general belief that the theory carries
within it the seeds of its own destruction, and should be expected to break
down in the TeV energy domain.  A number of suggestions have been made to
generalize the ideas of the SM to circumvent its problems but yet reduce to it
at energies $\l$ 100 GeV (and hence maintain its successes).  These include
supersymmetry models, technicolor models, and models based on the existance of
a $t\bar{t}$ condensate.  Each of these are different ways of treating the
quadratic divergence of the Higgs boson self energy.

The simplest technicolor models appear to be inconsistent with current LEP data
\cite{1}, though more complicated models may be able to evade this problem.
The simplest models of $t\overline{t}$ condensates favor a rather heavy
$t$-quark
($m_t \r 200$~Gev) and a heavy Higgs boson ($m_H \r 300$~GeV) which, while not
excluded, is not favored by current LEP data \cite{2}.  Again more complicated
models may be able to modify these predictions.  On the other hand, there has
been the first indication of supersymmetric grand unification arising from the
precision measurement at LEP and elsewhere of the SM coupling constants
$\alpha_1 (M_Z)$, $\alpha_2 (M_Z)$, $\alpha_3 (M_Z)$ at scale $Q = M_Z$.  Using
the renormalization group equations, grand unification does not occur for the
SM, but does appear to take place for its supersymmetrized version at a Gut
scale of \cite{3} $M_G \simeq 10^{16}$~GeV.  This may, of course, be just a
numerical accident, but if one takes this result seriously, it suggests the
validity of the combined ideas of supersymmetry and grand unification.
Further, in order to combine supersymmetry and grand unification, in a
phenomenologically viable way, one needs to make use of local supersymmetry
i.e. supergravity.  It thus appears to be of interest to ask what other
phenomena one can use to test the validity of supergravity grand unified
models.

In these talks, we will survey various aspects of supersymmetry and
supergravity Gut models.  While it will not be possible to give here the
derivation of many of the basic results, the talks will be pedagogic in the
sense that enough will be included to give a working knowledge of this area of
particle physics, as well as some of the recent results.  While supergravity
may be viewed to arise as a consequence of string theory, we won't impose any
particular string assumptions on the analysis.  The natural scale for
superstring theory is the Planck scale

\begin{equation}
M_{P\ell} = (8\pi G_N)^{-1/2} = 2.4 \times 10^{18}~GeV \label{1.1}
\end{equation} 

\noindent
Since $M_G/M_{P\ell} \approx 10^{-2}$ it may be that string effects are small
corrections to supergravity Gut models, and that such models are thus
reasonably self contained and represent another ``way station'' on the road to
a more fundamental theory.

\vglue 0.5cm
{\elevenbf \noindent 2.  What's Wrong with the Standard Model?}
\vglue 0.4cm

The Standard Model (SM) with three generations of quarks and leptons is based
on the gauge group $SU(3)_C \times SU(2)_L \times U(1)_Y$ where $C =$ color, $L
=$ left and $Y =$ hypercharge.  Quarks and leptons are left-handed doublets and
right-handed singlets:

\begin{equation}
q_i^{\alpha} = (u_i, d_i)_L;~u_{iR}, d_{iR};~l_i^{\alpha} = (\nu_i,
e_i)_L;~e_{iR} \label{2.1}
\end{equation} 

\noindent
where $i = 1, 2, 3$ is the generation index and $\alpha = 1, 2$ is the
$SU(2)_L$ doublet index.  In addition there is a Higgs doublet:

\begin{equation}
H^{\alpha} = (H^{(+)}, H^{(0)});~H_{\alpha} \equiv \varepsilon_{\alpha\beta}
H^{\beta} \label{2.2}
\end{equation} 

\noindent
with $\varepsilon_{\alpha\beta} = -
\varepsilon_{\beta\alpha}$~and~$\varepsilon_{12} = + 1$.

The dynamics of the SM consists of the gauge interactions, the Yukawa coupling
and the Higgs potential.  The first are constructed in the usual way:

\begin{equation}
\partial_{\mu} \rightarrow \partial_{\mu} - i [g_3 A_{\mu}^a T_C^a + g_2
B_{\mu}^m T_L^m + g^\prime B_{\mu} (Y/2)] \label{2.3}
\end{equation} 

\noindent
where $T_C^a = (\lambda^a/2;~0)$ for (quarks; leptons, Higgs), where  the
$\lambda^a$, $a = 1 \cdots 8$ are Gell-Mann matrices: $T_L^m = (\tau^m/2;~0)$
for $SU(2)_L$ (doublets; singlets), where the $\tau^m$ are Pauli matrices; and
$Y$ is defined by $Q = T_L^3 + Y/2$.  The $A_{\mu}^a (x)$, $B_{\mu}^m
(x)$~and~$B_{\mu}$ are the color gluons, and the $SU(2)_L$ and hypercharge
gauge bosons, and $g_3$, $g_2$~and~$g^\prime$ are their corresponding coupling
constants.

The Yukawa interactions are specified by the potential

\begin{equation}
V_Y = \lambda_{ij}^{(u)} H_{\alpha} \bar{u}_{iR} q_j^{\alpha} +
\lambda_{ij}^{(d)} H^{\alpha} \overline{q}_i^{\alpha} d_{Rj} +
\lambda_{ij}^{(e)}
H^{\alpha} \overline{l}_i^{\alpha} e_{Rj} \label{2.4}
\end{equation} 

\noindent
where $\lambda_{ij}^{(u,d,e)}$ are the Yukawa coupling constant.  The Higgs
potential is given by

\begin{equation}
V_H = - m^2 H^{\alpha \dagger} H^{\alpha} + \lambda (H^{\alpha \dagger}
H^{\alpha})^2;~m^2,
\lambda > 0 \label{2.5}
\end{equation} 

\noindent
The Higgs potential gives rise to a spontaneous breaking of $SU(2) \times U(1)$
due to the peculiar sign chosen in the mass term.  One finds, inserting
$H^{(0)} = (v + H^{(0)'})/\sqrt{2}$, where $v = \sqrt{2} <H^{(0)}>$, that
minimizing $V_H$ yields $v = m^2/\sqrt{2} \lambda$ and the Higgs boson
$H^{(0)'}$ then possesses a positive tree level (mass)$^2$ of $m_H^2 =
2m^2 > 0$.  The Yukawa interactions then gives rise to masses for the quarks
and leptons, and by making unitary transformations on the $R$~and~$L$ fields
one may go to the mass diagonal states, which for the down quarks are
$d'_i = V_{ij} d_j$ where $V$ is the CKM matrix.  The current
experimental evaluations of $V_{ij}$ are given in Ref.~[4].  The gauge
interactions also give rise to $W^{\pm}$~and~$Z$ boson masses e.g. $M_W = g_2
v/2$ which implies $v = 246$~GeV.  Alternately one may write

\begin{equation}
M_W = \frac{g_2}{\sqrt{\lambda}}~\frac{m_H}{2\sqrt{2}} \label{2.6}
\end{equation} 

\noindent
showing that $m_H$ sets the electroweak mass scale.

The SM has passed a very large number of experimental tests and is currently in
excellent agreement with all the data.  However, from the theoretical side,
many aspects of the SM are unsatisfactory, leading one to suspect that one is
seeing the low energy manisfestation of some more fundamental theory.  We list
here some of these difficulties:

\begin{description}
\item (i) The SM has 19 adjustable parameters.  These are 3 lepton masses
$(m_e, m_{\mu}, m_{\tau})$; 6 quark masses $(m_{u_i}, m_{d_i}, i = 1, 2, 3)$; 3
coupling constants $\alpha_3 = g_3^2/4\pi$, $\alpha_2 = g_2^2/4\pi$, $\alpha_Y
=
g'^2/4\pi$; 4 parameters in the CKM matrix (3 angles, 1 CP violating
phase); 2 Higgs potential parameters $(\lambda, m^2)$; and the strong CP
violating parameter $(\theta_{QCD} {\tilde F}_a^{\mu\nu} F_{\mu\nu}^a)$.  One
must go to experiment to fit these \cite{4}.  (The current bounds on
$m_t$~and~$m_H$ are $m_t > 118$~GeV (CDF) and $m_H > 62.5$~GeV (combined LEP
analysis).  Clearly a fundamental theory is expected to have fewer arbitrary
parameters.

\item (ii) The breaking of $SU(2) \times U(1)$ is inserted by hand (by choosing
$- m^2$ in Eq.~(\ref{2.5}) instead of $+ m^2$), rather than being a consequence
of the theoretical principles of the model.  Thus while the SM can accommodate
spontaneous electroweak breaking, it does not explain its origin.

\item (iii) There is no real electroweak unification as the SM group is a
product group.  Also there is a mysterious symmetry between quarks and leptons:
both have $L$ particles (not anti-particles) as $SU(2)$ doublets and both obey
the relation $Q = T_3 + Y/2$ which relates quark and lepton charges.  This
suggests the existance of grand unification with $SU(3) \times SU(2) \times
U(1)$ embedded in a simple group [e.g. $SU(5)$~or~$O(10)$] with quarks and
leptons appearing in the same representation.

\item (iv) Perhaps the most serious issue is the ``gauge hierarchy'' problem.
The loop corrections to $m_H$ (an example is given in Fig.~\ref{1}) are
quadratically
divergent leading to a mass

\begin{equation}
m_H\,^2 = 2m^2 + c ({\tilde \alpha}/4\pi) \wedge^2 \label{2.7}
\end{equation} 

\begin{figure}[htb]
\vspace{2.0in}
\caption{One loop correction to Higgs mass from Higgs couplings to quarks.}
\label{1}
\end{figure}

\noindent
where ${\tilde \alpha}$ is a coupling constant, and $\wedge$ is the cutoff.
One may view $\wedge$ as the scale where new physics occurs (which cuts off the
quadratic divergence).  How high can $\wedge$ be?  As $\wedge$ increases, $m_H$
(which by Eq.~(\ref{2.6}) scales electroweak physics) increases, and hence the
electroweak scale is driven close to the large scale $\wedge$.  This is known
as the ``gauge hierarchy'' problem.  An alternate possibility is to choose
$m^2$
to cancel the large loop correction.  This then results in the ``fine tuning''
problem.  For example if $\wedge = M_G \sim 10^{15}$~GeV, one must fine tune
$m^2$ to 24 decimal places (!) and in fact troubles begin already in the TeV
region.  Thus one expects the SM to break down in the TeV domain, which is why
the SSC and LHC are expected to uncover new physics.
\end{description}

\vglue 0.5cm
{\elevenbf \noindent 3.  Global Supersymmetry}
\vglue 0.4cm

Supersymmetry attempts to shed light on items (ii), (iii) and (iv) above.  It
only makes a small amount of progress on item (i), which presumably needs a
much deeper theory to approach (such as string theory).  We discuss first
global supersymmetry.

Supersymmetry is a symmetry between bosons and fermions, i.e. it requires that
the number of bose and fermi helicity states in a multiplet be equal.  It was
first introduced on purely ``aesthetic'' grounds, that nature should be even
handed between boson and fermions.

One may realize supersymmetry (SUSY) algebraically in the following way
\cite{5}.  Let $Q_{\alpha}$ be a $L$ spinor obeying the following
anti-commutation relations \cite{6}:

\begin{equation}
\{Q_{\alpha}, Q_{\beta}^{\dagger}\} = - 2 (P_L \gamma^{\mu}
\gamma^0)_{\alpha\beta}
P_{\mu} \label{3.1a}
\end{equation} 

\begin{equation}
\{Q_{\alpha}, Q_{\beta}\} = [Q_{\alpha}, P_{\mu}] = 0 = [P_{\mu},
P_{\nu}] \label{3.1b}
\end{equation} 

\noindent
where $P^{\mu}$ is the energy momentum vector and $P_L = (1 - \gamma^5)/2$.
Eqs.~(\ref{3.1a},\ref{3.1b}) are referred to as a ``Graded Lie Algebra''.  One
has that $M^2 \equiv - P^{\mu} P_{\mu}$ is a Casimir operator and so all states
in a SUSY multiplet have the same mass.  One may now verify that each such
multiplet has an equal number bose and fermi states:  Define the Witten index
$(-1)^{N_f}$ where $N_f$ is the operator whose eigenvalues are number of
fermions in a state.  Since $Q_{\alpha}$ is a fermionic operator, it must
anti-commute with $(-1)^{N_f}$.  Hence if $Tr$ is the trace over the states of
a
multiplet we have

\begin{eqnarray}
Tr [(-1)^{N_f} \{Q_{\alpha}, Q_{\beta}^{\dagger}\}] & = & Tr [- Q_{\alpha}
(-1)^{N_f} Q_{\beta}^{\dagger} + (-1)^{N_f} Q_{\beta}^{\dagger}
Q_{\alpha}]\nonumber \\
& = & 0 \label{3.2}
\end{eqnarray} 

\noindent
But by Eq.~(\ref{3.1a}) this implies

\begin{equation}
O = Tr (-1)^{N_f} P_{\mu} = P'_{\mu} Tr (-1)^{N_f} \label{3.3}
\end{equation} 

\noindent
where $P'_{\mu}$ is the (common) eigenvalue of the multiplet.  Thus $Tr
(-1)^{N_f} = 0$~for~$P'_{\mu} \not= 0$ and hence the number fermi and bose
states must be equal when summed over the multiplet.

The above considerations would be an amusing toy were it not for two remarkable
theorems.  The first of these is the following \cite{7}:  Aside from an
irrelevant generalization \cite{8}, the only graded algebra for an $S$-matrix
constructed from a local relativistic quantum field theory is the supersymmetry
algebra.  Thus supersymmetry is unique i.e. it is the only graded extension of
Lorentz covariant field theory.

We mention now several special features of SUSY systems.  Taking the trace of
Eq.~(\ref{3.1a}) gives

\begin{equation}
P^0 \equiv H = \frac{1}{4} (QQ^{\dagger} + Q^{\dagger} Q) \geq 0 \label{3.4}
\end{equation} 

\noindent
i.e. the Hamiltonian is always a positive semi-definite operator.  If the
vacuum state is supersymmetric, i.e. $Q_{\alpha} |0> = 0 = Q_{\alpha}^+ |0>$
then the vacuum energy vanishes:  $E_{vac} = <0| H |0> \equiv 0$.  If there is
spontaneous breaking of supersymmetry, i.e. $Q_{\alpha} |0> \not=
0$~and~$Q_{\alpha}^+ |0> \not= 0$, then $E_{vac} = <0| H |0>~> 0$.  This
suggests that it may be difficult to break supersymmetry as the symmetric
vacuum always lies lower than the broken vacuum.  To break global SUSY one
must arrange it so that the symmetric vacuum does not exist (i.e. is not an
extrema of the effective potential).  In fact, for a wide class of systems it
can be shown that if SUSY doesn't break at the tree level, it doesn't break
with quantum corrections \cite{9}.

The simplest SUSY multiplets are the massless ones which consist of states of
spin $s$~and~$s + 1/2$.  For model building, we will need the following:

\vglue 0.2cm
{\elevenit\noindent (i) Chiral multiplet:  $(z (x), \chi (x))$}
\vglue 0.1cm

This consists of $z (x)$, a complex scalar field $(s = 0)$, and $\chi (x)$ a
$L$ Weyl spinor.  There are thus 2 bose and 2 fermi states in this multiplet.
These multiplets can be used to represent matter, since quarks and leptons are
represented by $L$ Weyl spinors.  The $z(x)$ fields are additional spin zero
fields needed for supersymmetry-the ``squarks'' and ``sleptons''.

\vglue 0.2cm
{\elevenit\noindent (ii) Vector multiplet:  $(V^{\mu} (x), \lambda (x))$}
\vglue 0.1cm

This consists of $V^{\mu} (x)$, a real vector field $(s = 1)$~and~$\lambda (x)$
a Majorana spinor $(s = 1/2)$.  Again there are 2 bose and 2 fermi states (a
massless vector boson has only 2 helicity states).  This multiplet can be used
to represent gauge bosons, the additional spinor $\lambda (x)$ being the
supersymmetric ``gaugino'' partner.

The dynamics of global SUSY consists of supersymmetrized gauge and Yukawa
interactions:

\noindent
Gauge Interactions:  Here there are three types of terms:  (a) the bose gauge
interactions obtained in the usual way by the replacement

\begin{equation}
\partial_{\mu} \rightarrow \partial_{\mu} - i \sum_i g_i V_{\mu i}^a
T_i^a \label{3.5a}
\end{equation} 

\noindent
where $V_{\mu i}^a, g_i$~and~$T_i^a$ are the gauge bosons, gauge coupling
constants and group representations for the sub-group $G_i$.  (The full
symmetry group may be a product group $G = \prod_i G_i$.); (b) the fermi
gaugino interactions with the chiral multiplets $(z_m (x), \chi_m (x))$,

\begin{equation}
{\cal L}_{\lambda} = - i \sqrt{2} \sum_{i,m} g_i \bar{\lambda}_i^a
z_m^{\dagger}
T_i^a \chi_m \label{3.5b}
\end{equation} 

\noindent
and (c) an additional bose contribution to the effective potential, the
``$D$ term'',

\begin{equation}
V_D = \frac{1}{2} \sum_{i,a} g_i^2 D_i^a D_i^a;~D_i^a = \sum_m
z_m^{\dagger} T_i^a z_m \label{3.5c}
\end{equation} 

\noindent
Together, the interactions of Eqs.~(\ref{3.5a},\ref{3.5b},\ref{3.5c}) are
supersymmetric and gauge
invariant.

\noindent
Yukawa Interactions:  These are governed by a superpotential $W (z_m)$.  (Note
$W$ is ``holomorphic'' in the sense it is a function of the $z (x)$ but not
the $z^{\dagger} (x)$.)  The bose Yukawa interactions are the ``$F$ term''
contribution
to the effective potential

\begin{equation}
V_F = \sum_m \Bigg | \frac{\partial W}{\partial z_m} \Bigg |^2, \label{3.6a}
\end{equation} 

\noindent
and the fermi Yukawa interactions are given by the Lagrangian term,

\begin{equation}
{\cal L}_Y = - \frac{1}{2} \sum_{m,n} \overline{\chi}_m^C \frac{\partial^2
W}{\partial z_m \partial z_n}~\chi_n + h.c. \label{3.6b}
\end{equation} 

\noindent
where $\chi^C$ means charge conjugate field.  Again
Eqs.~(\ref{3.6a},\ref{3.6b})
together are supersymmetric.

Thus, given a gauge invariant superpotential $W$ and the gauge group $G$,
Eqs.~(\ref{3.5a})-(\ref{3.6b}) are a unique set of supersymmetric gauge
invariant interactions.

The above statement of supersymmetric dynamics leads to the second remarkable
theorem of supersymmetry i.e. supersymmetric systems as described above have
no quadratic divergences.  More precisely, there is no renormalization at all
for any of the couplings in the $F$ term (i.e. in $V_Y$).  The only infinities
of the theory are logarithmic infinities of wave function renormalizations and
gauge coupling constant renormalizations \cite{10}.  Thus, not only is SUSY the
unique graded extension of the SM \cite{7}, but it also eliminates the
quadratic
Higgs self mass divergence (which was the most serious theoretic disability of
the SM).

One can see how this comes about as follows.  Not only does the Higgs have the
usual interactions with quarks of the SM, but also, from $V_Y$, a squark
interaction is required to maintain supersymmetry.  This is shown in
Fig.~\ref{2}.
Both diagrams are quadratically divergent but with opposite sign.  With perfect
supersymmetry all divergences cancel!  When supersymmetry is broken, the quark
and squark masses are no longer degenerate and a logarithmic divergence
survives.  The cutoff $\wedge^2$ of Eq.~(\ref{2.7}) gets replaced, $\wedge^2
\rightarrow (m_{{\tilde q}}^2 - m_q^2) ln (\wedge^2/m_{{\tilde q}}^2)$.  To
avoid fine tuning we now require $m_{{\tilde q}}~\l~1$~TeV.  Thus we expect the
new SUSY particles to be within striking distance of current and planned
accelerators.

\begin{figure}[htb]
\vspace{2.0in}
\caption{Higgs one loop corrections in supersymmetric models.  Compare with the
Standard Model diagram of Fig.~1.}
\label{2}
\end{figure}

\vglue 0.5cm
{\elevenbf \noindent 4.  Minimal Supersymmetric Standard Model (MSSM)}
\vglue 0.4cm

The MSSM is the simplest supersymmetric extension of the Standard Model.  One
promotes each particle in the SM to either a chiral or vector supermultiplet.
The particle
content is given in Table~1.  Unique to the MSSM is the appearance now of two
Higgs doublets,
$H_1$~and~$H_2$.  Higgs doublets must come in pairs in SUSY models in order to
cancel anomalies.
Thus the Higgs boson has become a chiral multiplet, implying the existance now
of the fermion
Higgsinos, and hence to the danger of possible anomalies.  (Anomalies only
arise in fermionic
triangle loops and so the SM does not have this problem.)  A necessary
condition to cancel
anomalies is that $\sum Y_i = 0$ where $Y$ is the hypercharge.  One sees that
$Y_{H_1} = -
1$~and~$Y_{H_2} = + 1$ so that the anomaly cancellation condition is obeyed
when both Higgs
doublets are present.  (Anomalies in the quark and lepton sector of course
cancel by the usual
GIM mechanism).

\begin{table}[htb]
\baselineskip=12pt
\vglue 0.4cm
\begin{center}
\begin{tabular}{cc}\hline\hline
 &  \\
\multicolumn{2}{c}{Vector Multiplets} \\
$j = 1$ & $j = \frac{1}{2}$ \\ \hline
$g_{\mu}^a (x),~a = 1 \cdots 8$ & $\lambda^a (x),~a = 1 \cdots 8$ \\
gluons $(g)$ & gluinos $({\tilde g})$ \\
$B_{\mu}^{\alpha} (x),~B_{\mu}^Y (x),~\alpha = 1,2,3$ & $\lambda^{\alpha}
(x),~\lambda^Y (x),~\alpha = 1,2,3$ \\
$SU(2)_L \times U(1)_Y$ gauge bosons & $SU(2)_L \times U(1)_Y$ gauginos \\
\hline\hline
 &  \\
\multicolumn{2}{c}{Chiral Multiplets} \\
$j = \frac{1}{2}$ & $j = 0$ \\ \hline
$(u_{iL},~d_{iL});~u_{iR},~d_{iR}$ & $({\tilde u}_{iL},~{\tilde
d}_{iL});~{\tilde u}_{iR},~{\tilde d}_{iR}$ \\
quarks ($i =$~generation) & squarks \\
$(\nu_{iL},~e_{iL});~e_{iR}$ & $({\tilde \nu}_{iL},~{\tilde e}_{iL});~{\tilde
e}_{iR}$ \\
leptons & sleptons \\
${\tilde H}_1 = ({\tilde H}_1^0,~{\tilde H}_1^-);~{\tilde H}_2 = ({\tilde
H}_2^+,~{\tilde H}_2^0)$ & $H_1 = (H_1^0,~H_1^-);~H_2 = (H_2^+,~H_2^0)$ \\
Higgsinos & Higgs bosons \\ \hline\hline
\end{tabular}
\end{center}
\noindent
Table~1.  Particle content of the MSSM.
\end{table}
\vglue 0.5cm

The Yukawa couplings are obtained from the superpotential.  From
Eqs.~\linebreak (\ref{3.6a},\ref{3.6b}), renormalizability and gauge invariance
imply that $W$ is at
most a sum of quadratic and cubic terms,

\begin{equation}
W = W^{(2)} + W^{(3)}, \label{4.1}
\end{equation} 

\noindent
and the general form that preserves baryon number $(B)$ and lepton number
$(L)$ and is $SU(3)_C \times SU(2)_L \times U(1)_Y$ invariant is \cite{11}

\begin{equation}
W^{(2)} = \mu H_1^{\alpha} H_{2\alpha} \label{4.2a}
\end{equation} 

\begin{eqnarray}
W^{(3)} & = & \lambda_{ij}^{(u)} {\tilde q}_i^{\alpha} H_{2\alpha} {\tilde
u}_j^C + \lambda_{ij}^{(d)} {\tilde q}_i^{\alpha} H_{1 \alpha} {\tilde d}_j^C
\nonumber \\
& + & \lambda_{ij}^{(e)} {\tilde \ell}_i^{\alpha} H_{1 \alpha} {\tilde
e}_i^C \label{4.2b}
\end{eqnarray} 

\noindent
where $\lambda_{ij}^{(u,d,e)}$ are the Yukawa coupling constants as in
Eq.~(\ref{2.4}).  $\mu$ is a parameter with dimensions of mass.  Note that in
the
SM with only one Higgs doublet, $B$~and~$L$ are automatically conserved.  In
SUSY, however, this is not the case as $H_1$ has the same quantum numbers as
the
lepton doublets $\ell_i$.  Thus it is possible, for example, to violate $L$ by
adding to Eq.~(\ref{4.2a}) a term $\ell_i^{\alpha} H_{2\alpha}$ [and similarly
replace $H_1$ by $\ell_i$ in Eq.~(\ref{4.2b})].  We will not consider such
models
here.

$W^{(2)}$ is a general mixing term between the two Higgs doublets and is
scaled by the mass parameter $\mu$, while $W^{(3)}$ are the Yukawa couplings.
When $SU(2) \times U(1)$ breaks, one expects, in general, both neutral
components, $H_1^0$~and~$H_2^0$, to grow VeVs.  Thus one sees that $<H_2^0>$
will give rise to $u$-quark masses, and $<H_1^0>$ to $d$-quark and lepton
masses.  Further, one needs two distinct Higgs doublets to form the Yukawa
interactions of Eq.~(\ref{4.2b}) e.g. one could not use $H_1$ in the $u$-quark
term as it has the wrong hypercharge.  One would need to use $H_1^{\dagger}$ to
form an
$u$-quark Yukawa term.  But supersymmetry requires $W$ to be holomorphic, i.e.
a function of the scalar fields $z_m$ only and not of $z_m^{\dagger}$.  Hence a
term
with $H_1^{\dagger}$ is forbidden.  Thus, two Higgs doublets are needed in
supersymmetry both on fundamental grounds (to cancel anomalies) and on
phenomenological grounds (so that both $u$~and~$d$ quarks can grow masses), a
neat matching of rather disparat requirements.

After $SU(2) \times U(1)$ breaks, there is a mixing of Higgsinos and $SU(2)
\times U(1)$ gauginos.  Thus the $H_2$ part of Eq.~(\ref{3.5b}) for the $SU(2)$
gauginos is

\begin{eqnarray}
{\cal L}_{\lambda} & = & - i \sqrt{2} g_2 \bar{\lambda}^i H_2^{\alpha \dagger}
(\frac{\tau^i}{2})_{\alpha\beta} {\tilde H}_2^{\beta} + h.c. \nonumber \\
& & \rightarrow - i \sqrt{2} g_2 \bar{\lambda}^i
<H_2^{\alpha}>^{\textstyle{\ast}}
(\frac{\tau^i}{2})_{\alpha\beta} {\tilde H}_2^{\beta} \label{4.3}
\end{eqnarray} 

\noindent
Also, from Eq.~(\ref{3.6b}) $W^{(2)}$ gives a mixing between ${\tilde
H}_1$~and~${\tilde H}_2$:

\begin{eqnarray}
{\cal L}_Y & = & - \frac{1}{2}~\bar{\tilde H}_1^{\alpha C} \frac{\partial^2
W}{\partial H_1^{\alpha} H_{2\beta}}~{\tilde H}_{2\beta} + h.c. \nonumber \\
& & \rightarrow - \mu \overline{H}_1^{\alpha} H_{2\alpha} \label{4.4}
\end{eqnarray} 

\noindent
Thus one gets a mass matrix coupling Higgsinos and gauginos.  The mass diagonal
states are the 2 charginos (Winos) ${\tilde W}_i, i = 1,2$ (charged spin 1/2
Dirac fields) and 4 neutralinos (Zinos) $Z_i, i = 1 \cdots 4$ (neutral spin 1/2
Majorana fields).  (We label our states such that $m_i < m_j$ for $i < j$.).
The $W^{\pm}$~and~$Z^0$ bosons grow masses by the usual Higgs mechanism,
absorbing a charged Higgs field and a hermitian neutral Higgs field from the
$H_1$~and~$H_2$ doublets.  Thus there is left 3 neutral and one charged Higgs
bosons which we denote as follows:  $h^0$~and~$H^0$ (CP even states), $A^0$ (CP
odd state) and $H^{\pm}$ (charged state).  The $h^0$ is defined to be the Higgs
boson that most closely resembles the SM Higgs.

In summary then, the MSSM implies the existance of 32 SUSY particles (over and
above the usual quarks and leptons of the SM):  12 squarks, 9 sleptons, 2
Winos, 4 Zinos, 1 gluino and 4 Higgs bosons.

There still remains, however, the problem of supersymmetry breaking.  In a
supersymmetric multiplet, the bose and fermi states all have the same mass.
Thus, for example, one a priori expects the squarks to be degenerate with the
quarks.  However, no light bosons of this type exist experimentally, which must
mean that supersymmetry is a broken symmetry if it is to have any validity.
The only satisfactory way of breaking a symmetry is by spontaneous breaking.
As
we saw, however, supersymmetry is very resistent to breaking, and no physically
acceptable way of producing spontaneous breaking is known for global
supersymmetry.  One is thus reduced to adding, on purely phenomenological
grounds, symmetry breaking terms in order to construct a viable model.
However, one must do this in a fashion that still maintains the cancellation of
the quadratic divergences.  Such type of SUSY symmetry breaking is called
``soft breaking'' and only a limited number of symmetry breaking terms will do
this \cite{12}.

The general form of soft breaking terms that one can add to the
supersymmetrized Standard Model is a contribution to the effective potential of
the form

\begin{eqnarray}
V_{SB} & = & m_{ab}^2 z_a z_b^{\dagger} + [A_{ij}^{(u)} \lambda_{ij}^{(u)} q_i
H_2
u_j^C + A_{ij}^{(d)} \lambda_{ij}^{(d)} q_i H_1 d_j^C \nonumber \\
& + & A_{ij}^{(e)} \lambda_{ij}^{(e)} \ell_i H_1 e_j^C + B \mu H_1 H_2 +
h.c.] \label{4.5a}
\end{eqnarray} 

\noindent
and a gaugino mass term

\begin{equation}
{\cal L}_{mass}^{\lambda} = - {\tilde m}_i \bar{\lambda}^i \lambda^i
\label{4.5b}
\end{equation} 

\noindent
Here $\{z_a\}$ are the set of scalar fields of the chiral multiplets and
$\{\lambda^i\}$ are all the gauginos.  Thus $m_{ab}^2$ is the scalar mass
matrix, $A_{ij}^{(u,d,e)}$~and~$B$ are soft breaking constants (sometimes
called ``Polonyi'' constants) and ${\tilde m}_i$ are the gaugino Majorana
masses \cite{13}.  This general form of the soft breaking terms then depends on
137 parameters (or 87 if one assumes them all real)!  Many, of course, can be
eliminated on purely phenomenological grounds, but there would still be too
many for the model to have much predictive value.  Instead, a limited number
are assumed to exist in the so called ``Minimal Supersymmetric Standard Model''
(MSSM) which we now define by the following conditions:

\begin{description}
\item (i) The particle content is that of the supersymmetrized Standard Model
with one pair of Higgs doublets (no extra ``exotic'' particles).

\item (ii) All squarks (except perhaps the $t$-squarks) are degenerate.  All
sleptons are degenerate.

\item (iii) Gaugino masses ${\tilde m}_i,~i = 1,2,3$ for $U(1)_Y$, $SU(2)_L$,
$SU(3)_C$ obey
\end{description}

\begin{equation}
{\tilde m}_Y:~{\tilde m}_2:~{\tilde m}_3 =
\alpha_Y:~\alpha_2:~\alpha_3 \label{4.6}
\end{equation} 

\noindent
These assumptions greatly reduce the number of free parameters and make the
theory phenomenologically useful.  Further, as we will see later, the MSSM is
an approximation to the low energy limit of supergravity Gut models (and in
fact was constructed after the invention of the supergravity models).

\vglue 0.5cm
{\elevenbf \noindent 5.  Experimental Bounds on SUSY Masses}
\vglue 0.4cm

Experimental searches for SUSY particles have been on going for a decade now.
Current bounds come from the Tevatron (CDF) and from the absence of SUSY
particles in $Z$ decays at LEP.  We list these bounds now for reference.  All
analysis of experimental data has been done with the framework of the MSSM.

Lower bounds on the gluino mass $m_{{\tilde g}}$ and on the squark mass
$m_{{\tilde q}}$ are difficult to state as the bounds on the two particles are
correlated and the bounds also depend on the parameters $\mu$~and~$\tan \beta
\equiv <H_2>/<H_1>$.  At present, a full analysis of the CDF data has not been
made but roughly one may say $m_{{\tilde q}}$, $m_{{\tilde g}}~\r~100$~GeV if
$m_{{\tilde g}}~\l~400$~GeV.  Fig.~\ref{3} shows the parameter choice that has
been
analysed \cite{14}.  The excluded regions may change as $\mu$~and~$\tan \beta$
are varied.

\begin{figure}[htb]
\vspace{3.0in}
\caption{Excluded region in $m_{{\tilde g}} - m_{{\tilde q}}$ for $\mu = -
250$~GeV, $\tan \beta = 2$ [CDF].}
\label{3}
\end{figure}

Bounds on the other SUSY particles come from LEP \cite{15}.  Since SUSY
particles
are created in pairs, the absence of them in $Z$ decay generally gives a lower
bound of approximately $M_Z/2 \simeq 45$~GeV (unless the coupling to the $Z$ is
anomalously small).  One finds, in fact for the charged sleptons that
$m_{{\tilde e}}$, $m_{{\tilde \mu}}$, $m_{{\tilde \tau}} > 45$~GeV and
$m_{{\tilde \nu}}
> 42$~GeV.  The charginos and neutralinos are bounded by $m_{{\tilde W}_1} >
45$~GeV, and $m_{{\tilde Z}_1} > 20$~GeV and $m_{{\tilde Z}_2}
> 45$~GeV for $\tan \beta > 3$.  The Higgs boson bounds are $m_h > 43$~GeV,
$m_A > 20-44$~GeV, $m_{H^{\pm}} > 42$~GeV.  However, if $m_A$ is large $(m_A^2
\gg M_Z^2)$, the $h$ boson couplings are similar to the Standard Model Higgs
and the bound rises to $m_h > 62.5$~GeV.

One sees that if we expect the new SUSY particles to inhabit a domain from
somewhat below $M_Z$ up to about 1 TeV, the parameter space has not yet been
much explored, and it is not too surprising that SUSY particles have not yet
been discovered.  In the near future, some of these bounds can be increased.
Thus LEP200, with maximum luminosity and energy, should be able to detect the
$h$
boson with mass up to 95 GeV, and the ${\tilde W}_1$ up to 100 GeV.  The
Tevatron, with a data sample of 100 pb$^{-1}$, could detect squarks (except the
${\tilde t}_1$) up to about 200 GeV, the ${\tilde t}_1$ up to about 100 GeV and
the ${\tilde W}_1$ up to about 70 GeV (and perhaps higher in some models).  The
LHC and SSC will be able to detect the gluino and squarks up to 2 TeV, as well
as be able to see the lighter Wino, Zinos and $h$ boson.

\vglue 0.5cm
{\elevenbf \noindent 6.  Unification of Couplings}
\vglue 0.4cm

About three years ago it became possible to test the question of whether there
is unification of the $SU(3)_C$, $SU(2)_L$, and $U(1)_Y$ coupling constants at
a high mass scale.  This was a result of accurate measurements of $\alpha_1$,
$\alpha_2$~and~$\alpha_3$ at the $Z$ mass scale.  It is convenient to use the
fine structure constant $\alpha (M_Z)$~and~$(sm^2 \theta_W)_{\overline{ms}}$ as
input to determine $\alpha_1 \equiv 5\alpha/3 \cos^2 \theta_W$~and~$\alpha_2 =
\alpha/\sin^2 \theta_W$.  Current determinations are \cite{16,17}

\begin{equation}
\alpha^{-1} (M_Z) = 127.9 \pm 0.1 \label{6.1}
\end{equation} 

\begin{equation}
(\sin^2 \theta_W)_{\overline{ms}} = 0.2328 \pm 0.0007 \label{6.2}
\end{equation} 

\noindent
Note that $\alpha (M_Z)$ is much less accurately known then its value at the
Thompson limit, $\alpha^{-1} (m_e) = 137.0359895(61)$.  This is due to lack
of data in the (1-10) GeV region in $e^+e^-$ scattering needed to accurately
run $\alpha$ from $m_e$ to $M_Z$.  In fact, the error in $\alpha (M_Z)$ is the
major error arising in high precision tests of the electroweak sector of the
Standard Model.  One finds from Eqs.~(\ref{6.1}), (\ref{6.2}) for $\alpha_{1,2}
(M_Z)$ then

\begin{equation}
\alpha_1 (M_Z) \equiv (5/3) \alpha_Y = 0.016985 \pm 0.000020 \label{6.3a}
\end{equation} 

\begin{equation}
\alpha_2 (M_Z) = 0.03358 \pm 0.00011 \label{6.3b}
\end{equation} 

\noindent
and the World average for $\alpha_3 (M_Z)$ is \cite{18}

\begin{equation}
\alpha_3 (M_Z) = 0.118 \pm 0.007 \label{6.3c}
\end{equation} 

Using the Renormalization Group Equations (RGE) one can determine the
$\alpha_i$ at any other mass scale and see whether or not they meet at some
high scale $\mu = M_G$.  The RGE to 2-loop order for $\alpha_i (\mu)$, $i =
1,2,3$ are \cite{19}

\begin{equation}
\mu \frac{d\alpha_i (\mu)}{d \mu} = - \frac{1}{2\pi}~[b_i + \frac{1}{4\pi}
\sum_j b_{ij} \alpha_j (\mu)]~\alpha_i^2 (\mu) \label{6.4}
\end{equation} 

\noindent
where for the Standard Model, $b_i$~and~$b_{ij}$ are

\begin{equation}
b_i = (0, -22/3, -11) + N_F (4/3, 4/3, 4/3) + N_H (1/10, 1/6, 0) \label{6.5a}
\end{equation} 

\begin{equation}
b_{ij} = \left( \begin{array}{ccc} 0&0&0 \\ 0&-136/3&0 \\ 0&0&-102 \end{array}
\right) + N_F \left( \begin{array}{ccc} 19/15&3/5&44/15 \\ 1/5&49/3&4 \\
11/30&3/2&76/3 \end{array} \right) + N_H \left( \begin{array}{ccc} 9/50&9/10&0
\\ 3/10&13/6&0 \\ 0&0&0 \end{array} \right) \label{6.5b}
\end{equation} 

\noindent
For the MSSM one has

\begin{equation}
b_i = (0, -6, -9) + N_F (2, 2, 2) + N_H (3/10, 1/2, 0) \label{6.6a}
\end{equation} 

\begin{equation}
b_{ij} = \left( \begin{array}{ccc} 0&0&0 \\ 0&-24&0 \\ 0&0&-54 \end{array}
\right) + N_F \left( \begin{array}{ccc} 38/15&6/5&88/15 \\ 2/5&14&8 \\
11/15&3&68/3 \end{array} \right) + N_H \left( \begin{array}{ccc} 9/50&9/10&0
\\ 3/10&7/2&0 \\ 0&0&0 \end{array} \right) \label{6.6b}
\end{equation} 

In Eqs.~(\ref{6.5a}-\ref{6.6b}) $N_F$ is the number of families $(N_F =
3)$~and~$N_H =$~the number of Higgs doublets.  ($N_H = 1$ for the SM and $N_H =
2$ for the MSSM).  The difference between the $b_i$~and~$b_{ij}$ of the SM and
the MSSM is due, of course, to the different particle spectrum.

Using the experimental values of $\alpha_i (M_Z)$ as initial conditions, one
may check whether the $\alpha_i$ meet at $\mu = M_G$:

\begin{equation}
\alpha_i (M_G) = \alpha_G = \rm{Gut~scale~coupling~constant} \label{6.7}
\end{equation} 

\begin{figure}[htb]
\vspace{2.0in}
\caption{$\alpha_i^{-1} (\mu)$ for the Standard Model with 1 Higgs doublet.
The three coupling constants do not meet.}
\label{4a}
\end{figure}

\noindent
For the SM, grand unification fails by more than 7 std (Fig.~\ref{4a} Amaldi et
al [2]).  For
supersymmetry one may consider the simplest approximation of assuming all SUSY
particles are
degenerate at a common mass $M_S$.  Then acceptable unification does occur but
only for one pair of light Higgs doublets (Fig.~\ref{4b}, Fig.~\ref{4c} Amaldi
et al [2]).  A fit to
the data for the MSSM depends on the value of $\alpha_3 (M_Z)$.  One finds
$\alpha_G^{-1} = 25.4 \pm
1.7$~and

\begin{figure}[htb]
\vspace{2.0in}
\caption{$\alpha_i (\mu)$ for the MSSM with one pair of Higgs doublets showing
grand unification.}
\label{4b}
\end{figure}

\begin{figure}[htb]
\vspace{2.0in}
\caption{$\alpha_i (\mu)$ for the MSSM with two pairs of Higgs doublets.
While unification occurs, $M_G$ is too small to prevent rapid proton decay $p
\rightarrow e^+ \pi^0$~and~$M_S$ is too large to resolve the gauge hierarchy
problem.}
\label{4c}
\end{figure}

\begin{equation}
M_G \cong 10^{16.2 + 5.7 (\alpha_3/0.118 - 1)} \label{6.7a}
\end{equation} 

\begin{equation}
M_S \cong 10^{2.4 + 17.4 (1 - \alpha_3/0.118)} \label{6.7b}
\end{equation} 

\noindent
We see that there is an anti-correlation between $M_G$~and~$M_S$:

\begin{equation}
M_G \cong 10^{16.2}~[10^{2.4}/M_S~(GeV)]^{0.33} \label{6.8}
\end{equation} 

\noindent
i.e a larger $M_S$ gives a smaller $M_G$, and from Eq.~(\ref{6.7b}) a larger
$M_S$ corresponds to a smaller value of $\alpha_3$.

We note the following points concerning the above results:

\begin{description}
\item (i) The major experimental error in the analysis is due to the
experimental error in
$\alpha_3 (M_Z)$.  Last year there appeared to be the possibility of a
discrepency between the low energy evaluations of $\alpha_3 (M_Z)$ and the
high energy LEP results.  Recently, however, the LEP average has been lowered
to \cite{20} $\alpha_3 (M_Z) = 0.123 \pm 0.006$ while the deep inelastic
scattering evaluation is $\alpha_3 (M_Z) = 0.112 \pm 0.005$.  Thus the choice
of Eq.~(\ref{6.3c}) is in reasonable agreement with both types of
determinations.

\item (ii) It is $\alpha_1 \equiv (5/3) \alpha_Y$ that unifies with
$\alpha_2$~and~$\alpha_3$.  From low energy SUSY (e.g. the MSSM) this looks
mysterious, i.e. why a $5/3$ factor?  However, from Gut physics based on
$SU(5)$, $SO(10)$ and other
groups this is just what is expected, i.e. the unification should occur when
the
hypercharge is embedded in the grand unified group with a factor $5/3$.  This
implies that the unification of the couplings is not just a property of low
energy SUSY, but also requires information about Gut scale physics, i.e. the
MSSM by itself does not imply grand unification.

\item (iii) The SUSY masses are, of course, not all degenerate and one needs
to include SUSY mass splitting.  There are also similar mass thresholds at the
Gut scale and these can also modify the results.  However, one cannot discuss
this within the framework of the MSSM as one needs a Gut model.  We will come
back to this within the framework  of supergravity grand unification.

\item (iv) Unification of couplings does not appear to be good news for string
theory which also expects unification, but at the Planck scale $M_{P\ell} =
2.4 \times 10^{18}$~GeV.  Two possible explanations for this discrepency have
been proposed:  There  may be string threshold effects that modify the
unification point.  Alternately, the string model (vacuum state) may predict
the existance of ``exotic'' particles with masses (less than $M_G$) of just
the right values so that the (modified) RGE delay unification until
$M_{P\ell}$.  However, up to now, no string model that predicts the desired
properties has been constructed.
\end{description}

\vglue 0.5cm
{\elevenbf \noindent 7.  Local Supersymmetry (Supergravity)}
\vglue 0.4cm

In order to obtain a phenomenologically acceptable spontaneous breaking of
supersymmetry, it appears to be necessary to promote supersymmetry from a
global to a local symmetry (which theoretically is a natural thing to do).
This turns out to require that gravity be included into the analysis, i.e.
local supersymmetry is supergravity.  We give in this section a brief
discussion of the ideas of supergravity.

To see why local supersymmetry forces one to include gravity, consider the
simplest global SUSY system of a non-interacting chiral multiplet, $\{z (x),
\chi (x)\}$.  The Lagrangian is \cite{6}:

\begin{equation}
{\cal L} = - \partial_{\mu} z^{\dagger} \partial^{\mu} z - \overline{\chi} (-i)
\gamma^{\mu} \partial_{\mu} \chi \label{7.1}
\end{equation} 

\noindent
${\cal L}$ is invariant under the global SUSY transformation

\begin{equation}
\delta z (x) = \overline{\varepsilon} \chi (x);~\delta \chi (x) = -i
\gamma^{\mu}
[\partial_{\mu} z (x)] \varepsilon \label{7.2}
\end{equation} 

\noindent
where $\varepsilon$ is an infinitesimal anti-commuting constant spinor.
Suppose now we
replace $\varepsilon$ by $\varepsilon (x)$.  Now ${\cal L}$ is no longer
invariant i.e.

\begin{equation}
\delta {\cal L} = [\partial_{\mu} \bar{\varepsilon} (x)]~(\gamma^{\mu}
[\gamma^{\alpha} \partial_{\alpha} z (x)]~\chi) + {\rm divergence} \label{7.3}
\end{equation} 

\noindent
To compensate for this term, we proceed as usual by introducing a
compensating gauge field (the ``gravitino'')

\begin{equation}
\psi_{\alpha}^{\mu} (x) = {\rm spin}~3/2~{\rm field} \label{7.4}
\end{equation} 

\noindent
($\alpha =$~spinor index) with transformation rule

\begin{equation}
\delta \psi^{\mu} = \kappa^{-1} \partial^{\mu} \varepsilon (x);~[\kappa] =
({\rm mass})^{-1} \label{7.5}
\end{equation} 

\noindent
We add to ${\cal L}$ the interaction term

\begin{equation}
{\cal L}_1 = -\kappa \overline{\psi}_{\mu} \gamma^{\mu}~[\gamma^{\alpha}
\partial_{\alpha} z (x)]~\chi \label{7.6}
\end{equation} 

\noindent
which clearly cancels the $\partial_{\mu} \overline{\varepsilon} (x)$ term of
Eq.~(\ref{7.3}) under the combined transformations of Eqs.~(\ref{7.2}) and
(\ref{7.5}).  However, ${\cal L} + {\cal L}_1$ is still not supersymmteric,
since
one finds

\begin{equation}
\delta {\cal L}_0 + \delta {\cal L}_1 = \kappa \overline{\psi}_{\nu}
\gamma_{\mu} T^{\mu\nu} \varepsilon (x) \label{7.7}
\end{equation} 

\noindent
where $T^{\mu\nu}$ is the $z(x)$ stress tensor.  This term can be canceled
if we now add to the system a second field,

\begin{equation}
g_{\mu\nu} (x) = {\rm spin}~2~{\rm field} \label{7.8}
\end{equation} 

\noindent
with supersymmetry transformation

\begin{equation}
\delta g_{\mu\nu} = \kappa \overline{\psi}_{\mu} \gamma_{\nu} \varepsilon
(x) \label{7.9}
\end{equation} 

\noindent
and add the coupling

\begin{equation}
{\cal L}_2 = - g_{\mu\nu} T^{\mu\nu} \label{7.10}
\end{equation} 

\noindent
Thus the requirement of local supersymmetry leads to a gauge theory based on
the SUSY massless ``supergravity multiplet'':

\begin{eqnarray}
&&\{\psi^{\mu} (x),~g_{\mu\nu} (x)\} \nonumber \\
s &=& ~~~3/2, \qquad ~2 \label{7.11}
\end{eqnarray}

\noindent
$g_{\mu\nu} (x)$ is clearly the gravitational field (a massless spin 2
particle),  ${\cal L}_2$ being the gravitational coupling, and one recognizes
$\kappa^{-1}$ as the Planck mass:  $\kappa^{-1} = (8\pi G_N)^{-1/2} \equiv
M_{P\ell}$ (where $G_N$ is the Newtonian constant).

Next, let us consider the supergravity multiplet, neglecting coupling to
matter.  The supergravity Lagrangian is just the sum of the spin 2 plus spin
3/2 Lagrangians.  Thus introducing the vierbein $e_{\mu}^m (x)$ by

\begin{equation}
g_{\mu\nu} (x) = e_{\mu}^m (x) \eta_{mn} e_{\mu}^n (x) \label{7.12}
\end{equation} 

\noindent
the $N = 1$ supergravity Lagrangian can be written as \cite{21}

\begin{equation}
{\cal L}_{SG} = - \frac{1}{2} \kappa^{-2} e R (e_{\mu}^m, w_{\mu}^{mn}) -
\frac{1}{2} \varepsilon^{\mu\nu\rho\sigma} \overline{\psi}_{\mu} \gamma^5
\gamma_{\nu} D_{\rho} \psi_{\sigma} \label{7.13}
\end{equation} 

\noindent
where $e = \det~[e_{\mu}^m]$, $R$ is the curvature scalar, $w_{\mu}^{mn}$ is
the spin connection (determined by varying Eq.~(\ref{7.13}) with respect to
$w_{\mu}^{mn}$) and $D_{\rho} = \partial_{\rho} + \frac{1}{2} w_{\rho}^{mn}
\sigma_{mn}$, $\sigma_{mn} \equiv \frac{1}{4}~[\gamma_m, \gamma_n]$ is the
covariant derivative.  Eq.~(\ref{7.13}) is invariant under the local
supersymmetry transformations

\begin{equation}
\delta e_{\mu}^m = \frac{1}{2}~\kappa \overline{\varepsilon} (x) \gamma^m
\psi_{\mu};~\delta \psi_{\mu} = \kappa^{-1} D_{\mu} \varepsilon
(x) \label{7.14}
\end{equation} 

\noindent
and one finds $\delta w_{\mu}^{mn} = 0$.  Eq.~(\ref{7.14}) is just the
generalization of Eqs.~(\ref{7.5}) and (\ref{7.9}) to maintain the general
coordinate invariance required by the gravitational interactions.

\vglue 0.5cm
{\elevenbf \noindent 8.  Supergravity Coupling to Matter}
\vglue 0.4cm

Just as one couples Einstein gravity to matter, one needs to couple
supergravity to matter.  For physically interesting systems, we saw that matter
consists of a number of chiral multiplets (to represent quarks, leptons etc.)
and an arbitrary vector multiplet (to represent the gauge particles).  The
analysis of how to couple the matter multiplets to the supergravity multiplet
and maintain supergravity invariance is very complicated \cite{22,23}.  We give
here
only that part of the Lagrangian relative to Gut model building.

The general couplings of $N = 1$ supergravity depend upon three functions, the
superpotential ${\tilde W} (z_i)$, the Kahler potential $d (z_i,
z_i^{\dagger})$ and
the gauge kinetic function $f_{\alpha\beta} (z_i, z_i^{\dagger})$.  Here the
$\{z_i\}$
are the spin zero components of the left handed chiral multiplets.  $\{z_i,
\chi_i\}$, and $\alpha,~\beta$ are adjoint representation gauge indices.
${\tilde W}$, $d$~and~$f_{\alpha\beta}$ are hermitain, ${\tilde W}$~and~$d$ are
gauge singlets and $f_{\alpha\beta}$ is a gauge tensor.  Actually, ${\tilde
W}$~and~$d$ enter in a single combination

\begin{equation}
{\cal G} = -\kappa^2 d - \ell n [\kappa^6 W W^{\dagger}] \label{8.1}
\end{equation} 

\noindent
so the general supergravity Lagrangian really depends only on two functions
$f_{\alpha\beta}$~and~${\cal G}$.  From Eq.~(\ref{8.1}) one also sees that the
theory is invariant under a Kahler transformation

\begin{equation}
d \rightarrow d - f (z_i) - f^{\dagger} (z_i),~W \rightarrow e^{\kappa^2 f} W
\label{8.2}
\end{equation} 

\noindent
The ``Kahler metric'' is defined by

\begin{equation}
g_j^i = d_{,j}^{,i} \equiv \frac{\partial^2 d}{\partial z_i z_j^{\dagger}}~= -
\kappa^{-2} {\cal
G}_{,j}^{,i} \label{8.3}
\end{equation} 

\noindent
For the special case $d = \sum_i z_i z_i^{\dagger}$ one has $g_j^i =
\delta_j^i$
which is a flat Kahler metric.

The total Lagrangian is quite complicated containing many non-linear terms
scaled by factors of $(M_{P\ell})^{-n}$.  Most of these are negligibly small at
energies below the Planck scale.  There are five important terms that we list
now:

\vglue 0.2cm
{\elevenit \noindent (i) Effective Potential}
\vglue 0.1cm

The effective potential has the form

\begin{eqnarray}
V & = & -\kappa^{-4} e^{-{\cal G}}~[({\cal G}^{-1})_j^i {\cal G}_{,i} {\cal
G}^{,j} + 3] \nonumber \\
& + & \frac{g^2}{2}~[Re (f^{-1})_{\alpha\beta}]~D_{\alpha}
D_{\beta} \label{8.4a}
\end{eqnarray} 

\noindent
where $({\cal G}^{-1})_j^i$, $(f^{-1})_{\alpha\beta}$ are the matrix inverses
of ${\cal G}_{,j}^{,i}$, $f_{\alpha\beta}$~and~$g$ is the gauge coupling
constant.  The quantity $D_{\alpha}$ is given by

\begin{equation}
D_{\alpha} = - \kappa^{-2} {\cal G}^{,i} (T^{\alpha})_{ij} z_j \label{8.4b}
\end{equation} 

\noindent
where $T^{\alpha}$ is the group generator.  Re-expressing $V$ in terms of
${\tilde W}$~and~$d$ gives

\begin{equation}
V = e^{\kappa d}~[(d^{-1})_j^i (\frac{\partial {\tilde W}}{\partial z_i}~+
\kappa^2 d_{,i} {\tilde W}) (\frac{\partial {\tilde W}}{\partial z_j}~+
\kappa^2 d_{,j} {\tilde W})^{\dagger} - 3 \kappa^2 | {\tilde W} |^2]~+ V_D
\label{8.5}
\end{equation} 

\noindent
where $V_D$ is the last term of Eq.~(\ref{8.4a})

\vglue 0.2cm
{\elevenit \noindent (ii) Scalar Kinetic Energy}
\vglue 0.1cm

The spin zero kinetic energy is given by

\begin{equation}
- d_{,j}^{,i} (D^{\mu} z_i)~(D_{\mu} z_j)^{\dagger} \label{8.6}
\end{equation} 

\noindent
where $D_{\mu}$ is the usual gauge covariant derivative.  We will consider
Kahler metrics where the non-linear terms are scaled by $1/M_{P\ell} = \kappa$;

\begin{equation}
g_j^i = d_{,j}^{,i} = c_j^i + \kappa c_{jk}^i z_k + \cdots~. \label{8.7}
\end{equation} 

\noindent
After spontaneous breakings (of the Gut gauge group and supersymmetry) some of
the $z_k$ fields will grow VEVs.  One can write then $g_j^i = <g_j^i> +
g_j^{i'}$ where $<g_j^{i'}> = 0$.  One may then make unitary and scale
transformations on $z_i$, $z_i^{\dagger}$ which reduces $<g_j^i>$ to
$\delta_j^i$ and
hence the kinetic energy to canonical form i.e.

\begin{equation}
g_j^i \rightarrow \delta_j^i + \kappa c_{jk}^{i'}
z'_k;~<z'_{\kappa}> = 0 \label{8.8}
\end{equation} 

\noindent
The extra non-renormalizable terms in Eq.~(\ref{8.6}) are scaled by $\kappa =
1/M_{P\ell}$ and hence negligible below the Planck scale.

A similar analysis holds for the spinor kinetic energy term of the chiral
multiplet which has the form $- d_{,j}^{,i} \overline{\chi}^i \gamma^{\mu}
\hat{D}_{\mu} \chi^j$ where $\hat{D}_{\mu}$ is the gauge and gravitational
covariant derivative.  The same transformations used in the spin zero sector
also reduces the spinor kinetic energy to canonical form.

\vglue 0.2cm
{\elevenit \noindent (iii) Gauge Kinetic Energy}
\vglue 0.1cm

The gauge kinetic energy has the form

\begin{equation}
- \frac{1}{4}~(Re f_{\alpha\beta}) F_{\mu\nu}^{\alpha}
F^{\mu\nu\beta} \label{8.9}
\end{equation} 

\noindent
We consider $f_{\alpha\beta}$ whose non-linear parts are also scaled by
$1/M_{P\ell}$, i.e. $f_{\alpha\beta} = c_{\alpha\beta} + \kappa c_{\alpha\beta
i} z_i + \cdots$;
which after spontaneous breakings takes the form $c_{\alpha\beta}' + \kappa
c_{\alpha\beta i'} z'_i + \cdots$, $<z'_i> = 0$.  One may  now make orthorgonal
and
scale transformations on $F_{\mu\nu}^{\alpha}$ to reduce $f_{\alpha\beta}$ to
$f_{\alpha\beta} = \delta_{\alpha\beta} + \kappa c_{\alpha\beta
i}'' z''_i + \cdots$ and hence the kinetic energy to
canonical form.  The same transformations simultaneously reduces the gaugino
kinetic energy, $-\frac{1}{2}~$(Ref$_{\alpha\beta}) \overline{\lambda}^{\alpha}
\gamma^{\mu} \hat{D}_{\mu} \lambda^{\beta}$ to canonical form.

\vglue 0.2cm
{\elevenit \noindent (iv) Gaugino ``Mass'' Term}
\vglue 0.1cm

A term quadratic in the gaugino fields exists in the Lagrangian:

\begin{equation}
[\frac{1}{4}~\kappa^{-1} e^{-{\cal G}/2} ({\cal G}^{-1})_j^i {\cal G}^{,j}
f_{\alpha\beta, i}^{\dagger}]~\overline{\lambda}^{\alpha} \lambda^{\beta}
\label{8.10}
\end{equation} 

\noindent
When supersymmetry breaks, we will see that it is possible for the bracket of
Eq.~(\ref{8.10}) to grow a VEV and hence give rise to a gaugino mass.

\vglue 0.2cm
{\elevenit\noindent (v) Gravitino ``Mass'' Term}
\vglue 0.1cm

A term quadratic in the gravitino mass also exists in the Lagrangian,

\begin{equation}
[\kappa^{-1} e^{-{\cal G}/2}]~\overline{\psi}_{\mu} \sigma^{\mu\nu}
\psi_{\nu} \label{8.11}
\end{equation} 

\noindent
and if supersymmetry breaks, a VEV of the bracket of Eq.~(\ref{8.11}) gives
rise
to a gravitino mass.

The effective potential of Eq.~(\ref{8.5}) is the supergravity generalization
of
the global SUSY effective potential.  Thus if we consider the expansion of
Eq.~(\ref{8.8}) and take the limit $\kappa \rightarrow 0$ (i.e. $M_{P\ell}
\rightarrow \infty$), the first term of Eq.~(\ref{8.5}) reduces precisely to
$V_F$ of Eq.~(\ref{3.6a}) with the notational change $W \rightarrow {\tilde
W}$.
Similarly, $D_{\alpha}$ of Eq.~(\ref{8.4b}) reduces to

\begin{equation}
D_{\alpha} = (d^{,i} + \kappa^{-2} {\tilde W}^{,i}/{\tilde W})
(T^{\alpha})_{ij} z_j \label{8.12}
\end{equation} 

\noindent
However, ${\tilde W}^{ji} T_{ij}^{\alpha} z_j \equiv 0$ (since ${\tilde W}$
is a gauge singlet) and writing $d^{,i} = z_i^{\dagger} + \cdots$ one gets in
the
limit $\kappa \rightarrow 0$ that $D_{\alpha} = z_i^{\dagger} (T^{\alpha})_{ij}
z_j$
\cite{24}.  With the expansion of $f_{\alpha\beta} = \delta_{\alpha\beta} +
\cdots$ discussed below Eq.~(\ref{8.9}) we see that in the $\kappa \rightarrow
0$
limit $V_D$ of Eq.~(\ref{8.4a}) reduces precisely to Eq.~(\ref{3.5c}) of global
supersymmetry.  Thus global supersymmetry can be viewed as the limit of
supergravity when $M_{P\ell} \rightarrow \infty$.

However, Planck scale effects are crucial in that they allow spontaneous
breaking of supersymmetry to occur in a natural way.  Thus the full
supergravity effective potential of Eq.~(\ref{8.5}) differs from the global one
of $V_F + V_D$, [Eqs.~(\ref{3.5c}) and (\ref{3.6a})] by not being positive
definite.  Hence Eq.~(\ref{8.5}) can easily accommodate supersymmetry breaking.
A
simple example of this is the choice \cite{25} ${\tilde W} = m^2 (z + B)$
(where
$m^2$~and~$B$ are constants) with a flat Kahler potential $(d = \sum_i z_i
z_i^{\dagger})$.  Minimizing the potential of Eq.~(\ref{8.5}) yields the result

\begin{equation}
<z> = \kappa^{-1} a (\sqrt{2} - \sqrt{6});~a = \pm~1 \label{8.13}
\end{equation} 

\noindent
We see that $<z> = O (M_{P\ell})$ and so the Planck terms in $V$ are crucial
to achieve this result.  [One can also fine tune the cosmological constant to
zero (arrange that $V_{min} = 0$) by chosing $B = - \kappa^{-1} a (2 \sqrt{2}
- \sqrt{6})$.]  The existance of a non-zero $<z>$ is a signal of
supersymmetry breaking via a ``super-Higgs'' effect.  The chiral partner of
$z(x)$ ie. $\chi_z (x)$ is absorbed by the gravitino $\psi^{\mu} (x)$ which
then becomes massive with mass $m_{3/2}$.  (The extra two helicities of
$\chi_z$ then gives the $\psi^{\mu} (x)$ four fermionic helicity states, the
correct number for a hermitian massive spin 3/2 field.  The complex Higgs
field $z(x)$ has two bose degrees of freedom, which with the massless
graviton $g_{\mu\nu} (x)$ gives four bosonic states.)  From Eq.~(\ref{8.11})
one
has

\begin{equation}
m_{3/2} = \frac{1}{2}~\kappa^2 |<W (<z>) | e^{\frac{\kappa^2}{4}
<z>^2} \label{8.14}
\end{equation} 

\noindent
which shows that $m_{3/2} \approx \kappa m^2$.  We will see below that
supersymmetry breaking triggers $SU(2) \times U(1)$ breaking at the
electroweak scale so that $\kappa m^2 = O (M_Z)$, which implies $m = O
(10^{10}$~GeV) for this case.

An alternate way of breaking supersymmetry is via a gaugino condensate
\cite{26}, $m_S^3 \equiv <\lambda \gamma^0 \lambda> \not= 0$.  Then it turns
out that $m_{3/2} \sim \kappa^2 < \lambda \gamma^0 \lambda>$ which is of
electroweak size for $m_S = O (10^{12-13}$~GeV).  Such an effect requires
non-perturbative phenomena to occur, and hence it is difficult to make an
explicit calculation in this mechanism for supersymmetry breaking.

\vglue 0.5cm
{\elevenbf \noindent 9.  Supergravity Gut Models}
\vglue 0.4cm

We consider in this section the construction of grand unified models based on
supergravity \cite{27}.  The advantage of using local supersymmetry is that, as
we have seen, spontaneous breaking of supersymmetry is easily achieved as a
consequence of the supergravity interactions.  However, a new hierarchy
problem arises.  Since the super Higgs VEV $<z> = O (M_{P\ell})$, there is a
danger that the breaking of supergravity will communicate huge Planck size
masses to the physical particles.  The secret of preventing this is to allow
the super Higgs field to couple only very weakly to the physical matter
fields i.e. only gravitationally.  This can be accomplished by requiring:

\begin{description}
\item (i) The super Higgs field $z (x)$ be a gauge singlet.

\item (ii) ${\tilde W} (z_i) = W (z_a) + W_h (z)$ where $\{z_a\}$ are the
physical fields.
\end{description}

\noindent
Condition (i) guarentees that there are no gauge couplings between $z(x)$ and
the physical fields $\{z_a (x)\}$, while from Eq.~(\ref{8.5}) one sees that the
other couplings between $z(x)$~and~$z_a (x)$ are always scaled by powers of
$\kappa = 1/M_{P\ell}$ and hence gravitationally suppressed.  Thus (i) and
(ii) implies that supersymmetry breaking lives in a ``hidden'' sector,
screened from the physical sector by gravity.  (Such hidden sectors also exist
in
string theory.)

The discussion of the unification of the SM coupling constants suggests that
the SM group is a valid symmetry below the Gut scale $M_G$.  Thus as one
proceeds from $M_{P\ell}$ down to $M_G$, one expects the Gut gauge group $G$
of the physical sector to break to the SM at $\mu = M_G:~G \rightarrow
SU(3)_C \times SU(2)_L \times U(1)_Y$.  For the theory to be reasonable this
must happen spontaneously with some of the fields, $\{z_A (x)\}$ growing VEVs
and superheavy masses:

\begin{equation}
<z_A> = O (M_G);~M_{z_A} = O (M_G) \label{9.1}
\end{equation} 

One may now integrate out the superheavy fields $z_A$ and eliminate the super
Higgs fields to obtain an effective $SU(3) \times SU(2) \times U(1)$ theory
just below $M_G$.  A remarkable theorem then holds \cite{28}:

Consider a class of models which obey the following conditions:

\begin{description}
\item (i) There exists a ``hidden'' sector which is gauge singlet with
respect to the physical sector gauge group $G$ which breaks supersymmetry
(e.g. by a super Higgs or gaugino condensate).  The hidden sector
communicates only gravitationally with the physical sector.

\item (ii) There exists a Gut sector which breaks $G$ to the Standard Model
at scale $Q = M_G:  G \rightarrow SU(3)_C \times SU(2)_L \times U(1)_Y$.

\item (iii) After integrating out the superheavy fields and eliminating the
super Higgs fields, the only light particles remaining below the Gut scale
are those of the supersymmetrized Standard Model with one pair of light Higgs
doublets.

\item (iv) Any super Higgs couplings that appear in the Kahler potential are
generation independent.
\end{description}

\noindent
Then, for models where the non-linear parts of $f_{\alpha\beta}$~and~$d$ are
scaled by $\kappa \equiv 1/M_{P\ell}$, the renormalizable interactions below
$M_G$ (which equivalently arise in the limit $\kappa \rightarrow 0$) are
described by an effective superpotential with quadratic and cubic parts, $W =
W^{(2)} + W^{(3)}$, and effective potential $V$,

\begin{equation}
V = \{\sum_a |\frac{\partial W}{\partial z_a}|^2 + V_D\} + \{m_0^2 z_a
z_a^{\dagger}
+ (A_0 W^{(3)} + B_0 W^{(2)} + h.c.)\}, \label{9.2}
\end{equation} 

\noindent
and a universal gaugino mass term

\begin{equation}
{\cal L}_{mass}^{\lambda} = - m_{1/2} \overline{\lambda}^{\alpha}
\lambda^{\alpha} \label{9.3}
\end{equation} 

\noindent
If the model is $R$ parity invariant or alternately lepton number conserving,
then $W$ has the unique form (required by $SU(3) \times SU(2) \times U(1)$
invariance),

\begin{equation}
W = \mu_0 H_1 H_2 + [\lambda_{ij}^{(u)} q_i H_2 u_j^C + \lambda_{ij}^{(d)}
d_i H_1 d_j^C + \lambda_{ij}^{(e)} l_i H_1 d_j^C] \label{9.4}
\end{equation} 

\noindent
where $\lambda_{ij}^{(u,d,e)}$ are the Yukawa coupling constants [analogous
to those in Eq.~(\ref{2.4})] and $H_{1,2}$ the two light Higgs doublets.

Assumptions (i) - (iv) lead to a wide class of models.  Thus conditions (ii)
and (iii) are what is needed to achieve grand unification [i.e. the SM group
holds up to $M_G$ with the particle spectrum as stated in (iii)].  Condition
(i) is what is required maintain the gauge hierarchy and (i) and (iv)
guarantee the suppression of flavor changing neutral interactions.  Thus it
is difficult to remove any of these assumption and still maintain a
phenomenologically viable model.

Eqs.~(\ref{9.2}) and (\ref{9.4}) show that the effective theory below the Gut
scale
is a broken global SUSY theory ($V_D$ is the usual $D$ term) with four soft
breaking terms.  These are scaled by the mass parameters $m_0$ (universal
mass of the chiral multiplet spin zero fields), $m_{1/2}$ (universal gaugino
mass), $A_0$~and~$B_0$ (scaling factor of cubic and quadratic parts of the
superpotential).  These four constants are determined by the nature of the
hidden sector, i.e. they parameterize our ignorance of the hidden sector
where supersymmetry breaks (e.g. by a hidden sector superpotential $W_h (z)$
or by hidden sector gaugino condensates).  In addition there is a Higgs
mixing parameter $\mu_0$.

It is interesting to compare the simplicity of this result to global SUSY
theory.  There one added in soft breaking terms by hand, Eqs.~(\ref{4.5a}) and
(\ref{4.5b}), and
the most general form depended on 137 parameters.  Effectively the supergravity
theory determines 133 of these parameters!

We briefly indicate how the various soft breaking terms in Eqs.~(\ref{9.2}),
(\ref{9.3}) arise for the case of super Higgs breaking of supersymmetry.  Thus
from Eq.~(\ref{8.5}), the term $(d^{-1})_j^i [\kappa^2 d_{,i} {\tilde
W}]~[\kappa^2 d_{,j} {\tilde W}]^{\dagger}$ gives rise to $(\kappa^2 |{\tilde
W}|^2) z_i
z_i^{\dagger}$ and hence to $m_0^2 \sim$ \linebreak $(\kappa^2 < W_h>)^2$.
Similarly, cross terms in
Eq.~(\ref{8.5}) such as $(d^{-1})_j^i \kappa^2 d_{,i} {\tilde W} (\partial
{\tilde W}/\partial z_j)$ give rise to $z_i (\partial W^{(2,3)}/\partial z_i)
\kappa^2 <W_h>$ which yield $A_0$~and~$B_0$ type structures.  If
$f_{\alpha\beta} = \delta_{\alpha\beta} (1 + c \kappa z + \cdots)$ then from
Eq.~(\ref{8.10}), for $i = j = z$, the VEV of the bracket gives a term of form
$\kappa^3 <z W_h>$ and since $<z> \sim \kappa^{-1}$ one has $m_{1/2} \sim
\kappa^2 <W_h>$.  Finally we note that the Kahler potential can have a form
$d = z_i z_i^{\dagger} + f (z_i) + f^{\dagger} (z_i)$ and the most general
gauge invariant
and lepton number invariant quadratic piece for $f$ is $f (z_i) = c H_1 H_2$
where $c$ is dimensionless.  (Higher terms in $f(z_i)$ load to
non-renormalizable interactions scaled by $\kappa = 1/M_{P\ell}$.)  By the
Kahler transformation of Eq.~(\ref{8.2}) this term can be moved into the
superpotential yielding $W e^{\kappa^2 f} = W + c \kappa^2 H_1 H_2 W +
\cdots$ and hence a quadratic term $c \kappa^2 <W_h> H_1 H_2$ showing that
$\mu_0 \sim \kappa^2 <W_h>$.  Thus we see that all the parameters $m_0$,
$m_{1/2}$, $A_0$, $B_0$~and~$\mu_0$ are of the same general size, i.e.
$\kappa^2 <W_h>$, and we will see later that this mass is the electroweak
mass scale.

\vglue 0.5cm
{\elevenbf \noindent 10.  An Example:  $SU(5)$ Supergravity Gut Model}
\vglue 0.4cm

In this section we discuss an explicit example of a supergravity Gut model
based on the Gut group $G = SU(5)$.  Here each generation of quarks and
leptons are in the $10 + \overline{5}$ representations of $SU(5)$:

\begin{equation}
10 = M_i^{XY} = - M_i^{YX};~\overline{5} = \overline{M}_{iY};~i = 1,2,3~X, Y =
1
\cdots 5 \label{10.1}
\end{equation} 

\noindent
where $i$ is the generation index and $X,Y$ are group indices.  In addition,
the Higgs are in $5 + \overline{5}$ representations

\begin{equation}
5 = H_1^X;~\overline{5} = \overline{H}_{2X} \label{10.2}
\end{equation} 

\noindent
Thus if we decompose the $SU(5)$ labels into their $SU(3) \times SU(2)$
content, $X = (a, \alpha)~a = 1 \cdots 3$, $\alpha = 4,5$ then
$H_1^{\alpha}$~and~$\overline{H}_{2\alpha}$ are the $H_1, H_2$ doublets of the
Standard Model, and in addition there are two color triplets (3 and
$\overline{3}$) Higgs $H_1^a$, $\overline{H}_{2a}$.

The total superpotential has a Yukawa, a Gut and a hidden part:  ${\tilde W}
= W_Y + W_G + W_h$.  $W_h$ is chosen to break supersymmetry.  The $SU(5)$
invariant Yukawa interactions are

\begin{equation}
W_Y = \lambda_{ij}^1 \varepsilon_{XYZWU} H_1^X M_i^{YZ} M_j^{WU} +
\lambda_{ij}^2 \overline{H}_{2X} \overline{M}_{iY} M_j^{XY} \label{10.3}
\end{equation} 

\noindent
In writing Eq.~(\ref{10.3}) we have assumed invariance under the discrete
symmetry (matter parity) of $(M_i^{XY}, \overline{M}_{iX}) \rightarrow
(-M_i^{XY}, - \overline{M}_{iX})$, $(H_1^X, \overline{H}_{2X}) \rightarrow
(H_1^X, \overline{H}_{2X})$ to forbid the cubic interactions $\overline{M}_{iX}
\overline{M}_{jY} M_k^{XY}$ which would give rise to too rapid proton decay.
The
Gut sector (originally proposed in global SUSY  models \cite{29}) uses a 24
representation of $SU(5), \Sigma_Y^X$, to break $SU(5)$ to the SM.  The
form of $W_G$ is

\begin{eqnarray}
W_G &=& \lambda_1 [\frac{1}{3} Tr \Sigma^3 + \frac{1}{2} M Tr \Sigma^2]
\nonumber \\
&&\qquad + \lambda_2 \bar{H}_{2X} [\Sigma_Y^X + 2 M'
\delta_Y^X]~H_1^Y \label{10.4}
\end{eqnarray} 

\noindent
where $Tr$ is the trace over $SU(5)$ indices, and $M$, $M'$ are mass
parameters of $O (M_G)$.  In the following we will set $M' = M$.

One may now minimize the effective potential.  $W_h$ causes the breaking of
supersymmetry and gives rise to the soft breaking parameters of
Eq.~(\ref{9.4}).  Assuming these are much smaller than $M_G$, one finds upon
minimizing the effective potential of Eq.~(\ref{9.2}) that $\Sigma_Y^X$ grows a
VEV which to leading order is

\begin{equation}
{\rm diag} <\Sigma_Y^X> = M (2,2,2, -3,-3)~[1 + O
(\frac{m_{SB}}{M})] \label{10.5}
\end{equation} 

\noindent
where $m_{SB}$ is any of the soft breaking parameters $m_0$, $m_{1/2}$ etc.
(Since we will see that $m_{SB} = O (M_Z)$, characteristically $O (m_{SB}/M)
\approx 10^{-14}$ and hence negligible.  Eq.~(\ref{10.5}) clearly breaks
$SU(5)$
and preserves $SU(3) \times SU(2) \times U(1)$.  The $\Sigma_Y^X$,
$H_1^a$~and~$H_{2a}$ become superheavy with masses $O (M) \approx M_G$ while
the $SU(2)$ doublets $H_1^{\alpha}$, $\overline{H}_{2\alpha}$ remain light with
masses $O (m_{SB})$~and~$O (\mu_0)$.  Integrating out the superheavy fields
in $W_Y$~and~$W_G$ and eliminating the super Higgs (which effectively means
replacing it by its VEV) one finds that the renormalizable interactions are
governed by an effective superpotential which is precisely the form of
Eq.~(\ref{9.2}) [with effective potential and gaugino mass term of
Eqs.~(\ref{9.4}) and (\ref{9.3})].  Thus at the Gut scale, the theory has
precisely the form of the MSSM with $H_1 = H_1^{\alpha}$~and~$H_2 =
\overline{H}_{2\alpha}$ playing the role of the two light Higgs doublets but
with
four soft breaking terms and a Higgs mixing parameterized by

\begin{equation}
m_0, m_{1/2}, A_0, B_0~{\rm and}~\mu_0 \label{10.6}
\end{equation} 

There is one theoretically unpleasant aspect about the above model.  In
Eq.~(\ref{10.4}) we set $M' = M$, and it is this that kept the Higgs doublets
light (i.e. the doublet masses are proportional to $M' - M$.)  While the
no-renormalization theorem of supersymmetry guarantees that this condition is
maintained even with loop corrections, this is indeed an awkward fine tuning.
There are several ways of avoiding this problem:

\vglue 0.2cm
{\elevenit\noindent (i) Missing Partner Mechanism}
\vglue 0.1cm

By using the 75, 50 and $\overline{50}$ representations in $W_G$ to break
$SU(5)$, instead of the 24, one can prevent the Higgs doublets from growing
masses \cite{30}.  Recall that

\begin{eqnarray}
75 & = & \Sigma_{UW}^{XY}, \Sigma_{UY}^{XY} \equiv 0;~50 =
\Theta_{XYZ}^{UW},~\Theta_{XYW}^{UW} \equiv 0 \nonumber \\
\overline{50} & = & \overline{\Theta}_{UW}^{XYZ},~\overline{\Theta}_{UW}^{XYW}
\equiv 0 \label{10.7}
\end{eqnarray} 

\noindent
where subscripted indices are all anti-symmetric as are superscripted
indices.  Instead of Eq.~(\ref{10.4}) one can chose

\begin{eqnarray}
W_G &=& \lambda_1 \Theta_{XYZ}^{UW} \Sigma_{UW}^{XY} H_1^Z + \lambda_2
\overline{\Theta}_{UW}^{XYZ} \Sigma_{XY}^{UW} \overline{H}_{2Z} \nonumber \\
&&+ M \Theta_{XYZ}^{UW} \overline{\Theta}_{UW}^{XYZ} + f (\Sigma) \label{10.8}
\end{eqnarray} 

\noindent
where $f (\Sigma)$ is chosen so that $\Sigma_{UW}^{XY}$ grows a
non-vanishing VEV of $O (M)$.  Now the $SU(3)_C \times SU(2)_L$ content of
the 50 is

\begin{equation}
50 = (8,2) + (6,3) + (\overline{6},1) + (3,2) + (\overline{3},1) +
(1,1) \label{10.9}
\end{equation} 

\noindent
i.e. there is no doublet (1,2) term.  Thus when $\Sigma$ is replaced by its
VEV in Eq.~(\ref{10.8}) there is no piece of $\Theta$ to match up to form a
mass
term with the $H_1^{\alpha}$ doublet in the $\lambda_1$ term (and similarly
no piece of $\overline{\Theta}$ can form a mass term with the
$\overline{H}_{2\alpha}$ doublet in the $\lambda_2$ term).  Note that the 50
does contain a ($\overline{3}$,1) piece and the $\overline{50}$ contains a
(3,1)
piece and so the color triplets, $H_1^a$, $\overline{H}_{2a}$ do indeed become
superheavy.

\vglue 0.2cm
{\elevenit\noindent (ii) Higgs Doublets as Goldstone Bosons}
\vglue 0.1cm

An alternate procedure is to choose a form for $W_G$ with a larger global
symmetry which has the gauge group $G$ as a subgroup \cite{31}.  [For example,
in the case $G = SU(5)$, one may give $W_G$ a global $SU(6)$ symmetry by
extending $\Sigma_Y^X X, Y = 1 \cdots 6$ to be a 35 of $SU(6)$ and
similarly $H_1^X$, $\overline{H}_{2X}$ be 6 and $\overline{6}$ of $SU(6)$.]
When
the local $SU(5)$ breaks spontaneously, the global $SU(6)$ also breaks.  The
Higgs color triplets become superheavy as before, but the Higgs doublets are
the Goldstone bosons arising from the breaking of the global $SU(6)$.  Thus
they remain massless.  One in fact finds in this analysis that $M =
M'$ occurs automatically, which illustrates that one man's fine
tuning may be another man's group symmetry!

\vglue 0.2cm
{\elevenit\noindent (iii) Flipped $SU(5) \times U(1)$}
\vglue 0.1cm

A third method of keeping the Higgs doublets light is to change the group
symmetry to $SU(5) \times U(1)$ and flipping the embedding of the particle
spectrum \cite{32}.  Thus one uses a $10 + \overline{5} + 1$ of $SU(5)$ for
each
generation, but interchanges $u$~and~$d$ quarks and $e$~and~$\nu$ leptons,
the $e^C$ field then appears in the singlet representation with a right
handed neutrino replacing it in the 10 representation.  Instead of using a
24 to break $SU(5)$, one introduces 10 and $\overline{10}$ fields along with
the
usual $H_1$~and~$\overline{H}_2$.  The Gut Yukawa couplings are then $\lambda_1
10~10~H_1 + \lambda_2 \overline{10}~\overline{10}~\overline{H}_2$.  The
$SU(3)_C
\times SU(2)_L$ content of a 10 representation is $(\overline{3},1) + (3,2) +
(1,1)$
and so when the 10 grows a VEV one has a natural missing partner for the
doublets (e.g. no (1,2) piece in the 10 to form a mass term with the
$H_{1\alpha}$).  However, there
is a $(\overline{3},1)$ which combines with the $H_1^a$ to make the color
triplets superheavy.  The
drawback of this model is that it is not fully unified since $G$ is a product
group.  A possible
solution to this difficulty is discussed in Sec.~17.

\vglue 0.5cm
{\elevenbf \noindent 11.  Radiative Breaking of $SU(2) \times U(1)$}
\vglue 0.4cm

In the SM, the spontaneous breaking of $SU(2) \times U(1)$ is arranged by
hand by inserting in a negative (mass)$^2$ in the Higgs potential.  Thus the
SM accomodates the breaking of $SU(2) \times U(1)$ but does not deduce it's
existance from any of the basic principles of the theory.  One of the
remarkable features of supergravity grand unification is that it offers a
natural explanation of $SU(2) \times U(1)$ breaking arising from radiative
corrections which generate dynamically the required negative (mass)$^2$
\cite{33}.  Further, it is the spontaneous breaking of supersymmetry at the
Planck scale that triggers the breaking of $SU(2) \times U(1)$ at the
electroweak scale, a result that is a consequence of supergravity theory
\cite{34}.  Hence the two spontaneous breakings are unified.

We consider here the general supergravity model which is described at the
Gut scale by Eqs.~(\ref{9.2})-(\ref{9.4}).  Aside from the Yukawa coupling
constants
(which are the same as in the SM) this theory depends upon seven constants:
$m_{1/2}, m_0, A_0, B_0$; $\mu_0$; $\alpha_G, M_G$.  This might be compared
with the SM which depends  upon $m^2$, $\lambda$ (from the Higgs potential);
$\alpha_1, \alpha_2, \alpha_3$.  Thus one needs only two more parameters in
supergravity Gut models than in the SM.  (Also, $\alpha_G$~and~$M_G$ may be
viewed as having been ``measured'' by the unification analysis of Sec.~6.)
As we will see, radiative breaking of $SU(2) \times U(1)$ will allow us to
eliminate one more parameter, and so the general model depends on relatively
few parameters and therefore should have significant predictive power.

The Gut theory is initially defined at $M_G$, while experiment takes place
at the electroweak scale.  One may use the Renormalization Group Equations
(RGE) to connect these two domains.  For discussion of electroweak breaking
we need the part of the effective potential involving the Higgs fields
$H_1$~and~$H_2$.  The renormalization group improved Higgs potential is $V_H
= V_0 + \Delta V_1$ where $V_0$ is the tree part [which can be read off
Eqs.~(\ref{9.2}) and (\ref{9.4})] and $\Delta V_1$ is the one loop correction.
One
finds

\begin{eqnarray}
V_0 &=& m_1^2 |H_1|^2 + m_2^2 |H_2|^2 - m_3^2 (H_1 H_2 + h.c.) \nonumber \\
&&\qquad + \frac{1}{8}~({\rm g}_2^2 + g_Y^2)~(|H_1|^2 - |H_2|^2)^2
\label{11.1a}
\end{eqnarray} 

\noindent
and \cite{35}

\begin{equation}
\Delta V_1 = \frac{1}{64 \pi^2} \sum_a (-1)^{2s_a} n_a M_a^4 ln
[\frac{M_a^2}{e^{3/2}Q^2}~] \label{11.1b}
\end{equation} 

\noindent
All parameters, $m_i (t)$, ${\rm g}_2 (t)$, $g_Y (t)$ are ``running''
parameters at scale $Q$ where $t = ln [M_G^2/Q^2]$.  Thus

\begin{eqnarray}
m_i^2 (t) &=& m_{H_i}^2 (t) + \mu^2 (t);~i = 1,2 \nonumber \\
m_3^2 (t) &=& - B (t) \mu (t) \label{11.2}
\end{eqnarray} 

\noindent
with boundary conditions at $Q = M_G~(t = 0)$:

\begin{eqnarray}
m_i^2 (0) &=& m_0^2 + \mu_0^2,~i = 1,2;~m_3^2 (0) = - B_0 \mu_0 \nonumber \\
\alpha_2 (0) &=& \alpha_G = (5/3) \alpha_Y (0) \label{11.3}
\end{eqnarray} 

\noindent
In Eq.~(\ref{11.1b}), $M_a = M_a (v_1, v_2)$ is the tree level mass of
particle $a$ as function of the VEVs $v_i = <H_i>$, and
$s_a$~and~$n_a$ are the spin and number of helicity states of particle $a$.

The RGE are given at the end of this section.  They allow one to express
all the parameters in Eqs.~(\ref{11.1a},\ref{11.1b}) in terms of the Gut scale
parameters.  The
procedure one uses is as follows:  A specific model corresponds to a choice
of Gut scale parameters $m_0$, $m_{1/2}$, $A_0$, $B_0$, and $\mu_0$.  At
the Gut scale, all scalar particles have an initial positive (mass)$^2$ of
$m_0^2$.  One then integrates down in energy until a (mass)$^2$ turns
negative, signaling the breakdown of $SU(2) \times U(1)$ and a VEV growth
$v_i = <H_i> \not= 0$.  Let us first look at the tree potential $V_0$.
Here $SU(2) \times U(1)$ breaking implies that the determinant of the
(mass)$^2$ matrix be negative (so that one negative (mass)$^2$ eigenvalue
exists) i.e. from Eq.~(\ref{11.1a}),

\begin{equation}
{\cal D} \equiv m_1^2 m_2^2 - m_3^4 < 0 \label{11.4}
\end{equation} 

\noindent
Also, for a valid minimum, one requires that the potential be bounded from
below, which is satisfied if

\begin{equation}
{\cal L} \equiv m_1^2 + m_2^2 - 2 |m_3^2| > 0 \label{11.5}
\end{equation} 

\noindent
Turning now to the full potential, the minimization conditions, $\partial
V_H/\partial v_i = 0$ yields the relations

\begin{equation}
\frac{1}{2}~M_Z^2 = \frac{\mu_1^2 - \mu_2^2 \tan^2 \beta}{\tan^2 \beta -
1}~;~\sin 2\beta = \frac{2m_3^2}{\mu_1^2 + \mu_2^2} \label{11.6}
\end{equation} 

\noindent
where $\tan \beta \equiv v_2/v_1$, $\mu_i^2 = m_1^2 +
\Sigma_i$~and~$\Sigma_i$ is the loop correction

\begin{equation}
\Sigma_i = \frac{1}{32 \pi^2}~\Sigma_a (-1)^{2s_a} n_a M_a^2 ln
[M_a^2/e^{1/2} Q^2]~(\partial M_a^2/\partial v_i) \label{11.7}
\end{equation} 

\noindent
Since $|\sin 2 \beta| \leq 1$, the second equation of (\ref{11.6}) includes a
generalization of the stability condition Eq.~(\ref{11.5}) with loop
corrections.

Normally, one would use minimization of $V_H$ to solve for the two VEVs
$v_i = <H_i>$.  Instead, it is more convenient to use them to eliminate two
of the Gut parameters in terms of $\tan \beta$ and the other parameters.  A
convenient choice is to use the second equation of (\ref{11.6}) to eliminate
$B_0$ and the
first equation to eliminate $\mu_0$.  Since only $\mu^2$ enters into the
first equation, there are two branches depending on the sign of $\mu$ i.e.
$\mu > 0$~and~$\mu < 0$.  One then has that the low energy physics depends
only on four parameters:

\begin{equation}
m_0, m_{1/2}, A_0, \tan \beta = v_2/v_1 \label{11.8}
\end{equation} 

\noindent
Thus the 32 SUSY particle masses can be determined in terms of these four
parameters (and the as yet unknown $t$-quark mass $m_t$).

There are several fine points worth mentioning:

\begin{description}
\item (i) If we define $Q_0$ as the mass scale where ${\cal D} (Q)$ of
Eq.~(\ref{11.4}) vanishes and $Q_1$ the point where ${\cal L} (Q)$ vanishes, in
general one finds that $Q_0 > Q_1$.  Thus at the tree level, satisfactory
electroweak
breaking occurs when $Q_1 < Q < Q_0$.  In running the RGE from $M_G$ down to
$Q$, what
value should we choose for $Q$?  (Where should we stop the RG ``clock''?)  It
has been
shown that when the loop correction $\Delta V_1$ is included, the results are
approximately independent of $Q$ \cite{36}.  In the following we will set $Q =
M_Z$.
This is convenient since then $M_Z (Q = M_Z)$ is the physical (experimental)
$Z$ boson
mass.  [All other masses then are running masses at scale $Q = M_Z$ (and can
differ from
their physical values)].  Actually, matters are simpler.  In most of the
parameter space, there is a large amount of cancelation in $\Sigma_i$
and so the one loop correction is small \cite{37}.  Thus using $V_0$ alone
generally gives a good approximation, and the $Q$ dependence of results are
small.

\item (ii) In the analysis given above, we have tacitly assumed that it is
a Higgs (mass)$^2$ that turns negative as one runs the RGE from $M_G$ to
the electroweak scale.  It is possible that some other (mass)$^2$ might
turn negative first e.g. a squark or slepton.  This would lead to a
catastrophe that $SU(3)_C$ or electromagnetism breaks down.  A number of
necessary conditions for this not to happen have been established \cite{38}
(though sufficiency conditions are not known).  Throughout most of the
parameter space, the known conditions are satisfied, and it is $m_{H_2}^2$
which turns negative.  The characteristic situation is shown schematically
in Fig.~\ref{5}.
\end{description}

\begin{figure}[htb]
\vspace{2.75in}
\caption{Schematic diagram of running of masses from $M_G$ to electroweak
scale.  The heavy top quark bends $m_{H_2}^2$ downwards as $Q$ decreases.}
\label{5}
\end{figure}

The existance of solutions to Eqs.~(\ref{11.6}) is the necessary conditions
that
electroweak breaking actually occurs.  No satisfactory solutions will exist,
however, unless
three requirements are met:

\begin{enumerate}
\item At least one of the soft breaking parameters $m_0$, $m_{1/2}$, $A_0$,
$B_0$ are non-zero

\item $\mu_0$ is non-zero

\item $m_t$ is heavy $(m_t \r 90$~GeV)
\end{enumerate}

\noindent
Condition (1) shows in a real sense that the breaking of supersymmetry at the
Gut scale generates $SU(2) \times U(1)$ breaking at the electroweak scale for
if all the soft breaking parameters generated by supergravity vanished, there
would be no electroweak breaking.  Similarly $\mu_0$ (which is also generated
from supergravity interactions) must be non-zero and $m_t$ must be heavy since
it is the $t$-quark Yukawa coupling to $H_2$ that drives $m_{H_2}^2$ negative.
Thus there is a confluence of three items, two of them theoretical aspects of
supergravity, and one of them experimental that gives rise to electroweak
breaking.  The fact that the top quark must be heavy was one of the first
predictions of
supergravity Guts \cite{33}.

\vglue 0.2cm
{\elevenit\noindent Renormalization Group Equations}
\vglue 0.1cm

We list here for reference the one loop RGE.  For simplicity, we keep only the
large $t$-quark Yukawa couplings, which is a good approximation for $\tan \beta
< 10$.  Our notation follows mainly that of Iba\~nez et al \cite{33} (except
that
$\alpha_1 \equiv (5/3) \alpha_Y)$.  Unfortunately, a number of different sign
conventions for $t$, $A_t$~and~$\mu$ exist in the literature.  We use $t = ln
(M_G^2/Q^2)$.

\begin{description}
\item (i) Gauge couplings:

\begin{equation}
\frac{d {\tilde \alpha}_i}{d t}~= - b_i {\tilde \alpha}_i^2 (t);~b_i =
(33/5, 1, -3), {\tilde \alpha}_i = \alpha_i/4\pi \label{11.9}
\end{equation} 

with boundary conditions $\alpha_i (0) = \alpha_G$.

\item (ii) Gaugino masses:

\begin{equation}
\frac{d {\tilde m}_i}{dt}~= - b_i {\tilde \alpha}_i (t) {\tilde
m}_i (t);~{\tilde m}_i (0) = m_{1/2} \label{11.10}
\end{equation} 

\item (iii) Higgs masses:

\begin{equation}
\frac{d m_{H_1}^2}{dt}~= 3 {\tilde \alpha}_2 {\tilde m}_2^2 + (3/5)
\tilde{\alpha}_1 {\tilde m}_1^2 \label{11.11}
\end{equation} 

\begin{eqnarray}
\frac{d m_{H_2}^2}{dt}~&=& (3 {\tilde \alpha}_2 {\tilde m}_2^2 + (3/5) {\tilde
\alpha}_1 {\tilde m}_1^2) \nonumber \\
&&-3 Y_t (m_Q^2 + m_U^2 + m_{H_2}^2 + A_t^2) \label{11.12}
\end{eqnarray} 

with boundary conditions $m_{H_i} (0) = m_0$.  (Note that it is the $t$-quark
Yukawa couplings, $Y_t$, which causes $m_{H_2}^2$ to decrease as $t$
increases.)

\item (iv) $t$-quark Yukawa coupling:

\begin{equation}
\frac{d Y_t}{dt}~= (\frac{16}{3}~{\tilde \alpha}_3 + 3 {\tilde \alpha}_2 +
\frac{13}{15}~{\tilde \alpha}_1 - 6 Y_t) Y_t \label{11.13}
\end{equation} 

where $Y_t = \lambda_t^2/4\pi$~and~$\lambda_t (Q)$ is the $t$-quark Yukawa
coupling constant defined by $m_t = \lambda_t (m_t) <H_2>$.

\item (iv) $t$-quark $A$ parameter:

\begin{equation}
\frac{d A_t}{dt}~= (16/3) {\tilde \alpha}_3 {\tilde m}_3 + 3 {\tilde \alpha}_2
{\tilde m}_2 + (13/15) {\tilde \alpha}_1 {\tilde m}_1) - 6 Y_t
A_t \label{11.14}
\end{equation} 

where $A_t (0) = A_0$.

\item (v) $B$ parameter:

\begin{equation}
\frac{dB}{dt}~= (3 {\tilde \alpha}_2 {\tilde m}_2 + (3/5) {\tilde \alpha}_1
{\tilde m}_1) - 3 Y_t A_t \label{11.15}
\end{equation} 

where $B (0) = B_0$.

\item (vi) $\mu$ parameter:

\begin{equation}
\frac{d \mu^2}{dt}~= (3 {\tilde \alpha}_2 + (3/5) {\tilde \alpha}_1 - 3 Y_t)
\mu^2 \label{11.16}
\end{equation} 

where $\mu (0) = \mu_0$.  (Note that the sign of $\mu_0$ is undetermined by
the $\mu$ RGE.)

\item (vii) Squark masses $(i = 1,2)$:

\begin{eqnarray}
Q_i &\equiv& ({\tilde u}_{iL}, {\tilde d}_{iL});~U_i \equiv {\tilde u}_{Ri},
D_i \equiv {\tilde d}_{Ri} \nonumber \\
\frac{dm_{Q_i}^2}{dt}~&=&(16/3) {\tilde \alpha}_3 {\tilde m}_3^2 + 3 {\tilde
\alpha}_2 {\tilde m}_2^2 + (1/15) {\tilde \alpha}_1 {\tilde
m}_1^2 \label{11.17}
\end{eqnarray} 

\begin{equation}
\frac{d m_{U_i}^2}{dt}~= (16/3) {\tilde \alpha}_3 {\tilde m}_3^2 + (16/15)
{\tilde \alpha}_1 {\tilde m}_1^2 \label{11.18}
\end{equation} 

\begin{equation}
\frac{d m_{D_i}^2}{dt}~= (16/3) {\tilde \alpha}_3 {\tilde m}_3^2 + (4/15)
{\tilde \alpha}_1 {\tilde m}_1^2) \label{11.19}
\end{equation} 

where $m_{Q_i} (0) = m_{U_i} (0) = m_{D_i} (0) = m_0$.

\item (viii) Slepton masses $(i = 1,2,3)$

\begin{eqnarray}
L_i \equiv ({\tilde \nu}_{Li}, {\tilde e}_{Li});~{\tilde E}_i \equiv {\tilde
e}_{Ri} \nonumber \\
\frac{d m_{Li}^2}{dt}~= 3 \alpha_2 {\tilde m}_2^2 + (3/5) {\tilde \alpha}_1
{\tilde m}_1^2 \label{11.20}
\end{eqnarray} 

\begin{equation}
\frac{dm_{Ei}}{dt}~= (12/5) {\tilde \alpha}_1 {\tilde m}_1^2 \label{11.21}
\end{equation} 

where $m_{Li} (0) = m_{Ei} (0) = m_0$.  (Note that the slepton and first two
generations of squark (mass)$^2$ all increase with increasing $t$.

\item (ix) Squark masses $(i = 3)$

\begin{eqnarray}
Q &\equiv& ({\tilde u}_{3L}, {\tilde d}_{3L});~U \equiv {\tilde u}_{3R}, D
\equiv {\tilde d}_{3R} \nonumber \\
\frac{d m_Q^2}{dt}~&=& (16/3) {\tilde \alpha}_3 {\tilde m}_3^2 + 3 {\tilde
\alpha}_2 {\tilde m}_2^2 + (1/15) {\tilde \alpha}_1 {\tilde m}_1^2 \nonumber \\
&-& Y_t (m_{H_2}^2 + m_Q^2 + m_U^2 + A_t^2) \label{11.22}
\end{eqnarray} 

\begin{eqnarray}
\frac{d m_U^2}{dt}~&=& (16/3) {\tilde \alpha}_3 {\tilde m}_3^2 + (16/15)
{\tilde \alpha}_1 {\tilde m}_1^2 \nonumber \\
&-& 2 Y_t (m_{H_2}^2 + m_Q^2 + m_U^2 + A_t^2) \label{11.23}
\end{eqnarray} 

\begin{equation}
\frac{d m_D^2}{dt}~= (16/3) {\tilde \alpha}_3 {\tilde m}_3^2 + (4/15) {\tilde
\alpha}_1 {\tilde m}_1^2 \label{11.24}
\end{equation} 

where $m_Q (0) = m_U (0) = m_D (0) = m_0$.
\end{description}

The procedure of solving these equations is the following:  (1) Solve
(\ref{11.9})
for $\alpha_i$ and insert into (\ref{11.10}) to solve for ${\tilde m}_i$.  (2)
One
can then integrate (\ref{11.11}), (\ref{11.17})-(\ref{11.19}) for
$m_{H_1}^2$, $(m_{Q_i}^2, m_{U_i}^2, m_{D_i}^2, i = 1,2)$, (\ref{11.20}) and
(\ref{11.21}) for $(m_{Li}^2$, $m_{Ei}^2, i = 1,2,3)$ and (\ref{11.24}) and
(\ref{11.13}) for
$m_D^2$ and $Y_t$.  (3) Using $Y_t$ one can solve (\ref{11.14}) and
(\ref{11.16}) for
$A_t$~and~$\mu^2$ and then using $Y_t$~and~$A_t$ one determines $B$ from
(\ref{11.15}).
(4) The three remaining equations, (\ref{11.12}), (\ref{11.22}), and
(\ref{11.23}) are coupled and
must be solved simultaneously.

\vglue 0.5cm
{\elevenbf \noindent 12.  Expressions for SUSY Masses}
\vglue 0.4cm

The SUSY masses are quantities that are defined at the low energy electroweak
scale and are experimentally accessible to current and future accelerators.
Using the RGE one can express each SUSY mass in terms of the $4 + 1$
parameters which define the Gut model:  $m_0$, $m_{1/2}$, $A_0$, $\tan$,
$\beta$~and~$m_t$.  In
this section, we discuss the explicit formulae for each SUSY mass.

\vglue 0.2cm
{\elevenit\noindent (i) Gauginos}
\vglue 0.1cm

{}From Eqs.~(\ref{11.9}), (\ref{11.10}) one finds ${\tilde m}_i (t) = [\alpha_i
(t)/\alpha_G] m_{1/2}$ where $\alpha_i (t) = \alpha_G/(1 + \beta_i
t)$~and~$\beta_i =
b_i \alpha_G/4\pi$.  These relations then imply

\begin{equation}
{\tilde m}_1:~{\tilde m}_2:~{\tilde m}_3 =
\alpha_1:~\alpha_2:~\alpha_3 \label{12.1}
\end{equation} 

\noindent
a relation that is postulated in the MSSM.

\vglue 0.2cm
{\elevenit\noindent (ii) Squarks $(i = 1,2)$; Sleptons $(i = 1,2,3)$}
\vglue 0.1cm

The squark and slepton masses can be obtained from the mass matrix $m_{ab}^2 =$
\linebreak
$<\partial^2 V^{eff}/\partial z_a \partial z_b^{\dagger}>$ where $V^{eff}$ is
the low
energy renormalization group improved effective potential derived from
Eq.~(\ref{9.2}).  Using Eqs.~(\ref{11.17})-(\ref{11.21}) to relate the low
energy mass
parameters to the Gut scale parameters for the first two generations of squarks
and all the sleptons,
one finds

\begin{eqnarray}
m_{{\tilde u}_{iL}}^2 = m_0^2 + m_{u_i}^2 &+& {\tilde \alpha}_G [(8/3) f_3 +
(3/2) f_2 + (1/30) f_1] m_{1/2}^2 \nonumber \\
&+& (\frac{1}{2}~-\frac{2}{3}~\sin^2 \theta_W) M_Z^2 \cos 2\beta \label{12.2a}
\end{eqnarray} 

\begin{eqnarray}
m_{{\tilde d}_{iL}}^2 = m_0^2 + m_{d_i}^2 &+& {\tilde \alpha}_G [(8/3) f_3 +
(3/2) f_2 + (1/30) f_1] m_{1/2}^2 \nonumber \\
&+& (-\frac{1}{2}~+\frac{1}{3}~\sin^2 \theta_W) M_Z^2 \cos 2\beta \label{12.2b}
\end{eqnarray} 

\begin{eqnarray}
m_{{\tilde u}_{iR}}^2 = m_0^2 + m_{u_i}^2 &+& {\tilde \alpha}_G [(8/3) f_3 +
(8/15) f_1] m_{1/2}^2 \nonumber \\
&+& (2/3) \sin^2 \theta_W M_Z^2 \cos 2\beta \label{12.2c}
\end{eqnarray} 

\begin{eqnarray}
m_{{\tilde d}_{iR}}^2 = m_0^2 + m_{d_i}^2 &+& {\tilde \alpha}_G [(8/3) f_3 +
(2/15) f_1] m_{1/2}^2 \nonumber \\
&+& (-1/3) \sin^2 \theta_W M_Z^2 \cos 2\beta \label{12.2d}
\end{eqnarray} 

\noindent
where $m_{u_i}$, $m_{d_i} i = 1,2$ are quark masses, ${\tilde \alpha}_G =
\alpha_G/4\pi$~and~$f_k (t) = t (2 - \beta_k t)/$ \linebreak $(1 + \beta_k
t)^2$, $\beta_k =
(33/5, 1, -3)$.  For the sleptons one similarly has

\begin{eqnarray}
m_{{\tilde e}_{iL}}^2 = m_0^2 + m_{ei}^2 &+& {\tilde \alpha}_G [(3/2) f_2 +
(3/10) f_1] m_{1/2}^2 \nonumber \\
&+& (- \frac{1}{2}~+ \sin^2 \theta_W) M_Z^2 \cos 2\beta \label{12.3a}
\end{eqnarray} 

\begin{equation}
m_{{\tilde e}iR}^2 = m_0^2 + m_{e_i}^2 + {\tilde \alpha}_G (6/5) f_1 m_{1/2}^2
- \sin^2 \theta_W M_Z^2 \cos 2\beta \label{12.3b}
\end{equation} 

\begin{equation}
m_{{\tilde \nu}_{iL}}^2 = m_0^2 + {\tilde \alpha}_G [(3/2) f_2 + (3/10) f_1]
m_{1/2}^2 + (1/2) M_Z^2 \cos 2 \beta \label{12.3c}
\end{equation} 

\noindent
where $m_{ei}$, $i = 1,2,3$ are the charged lepton masses.  The quark and
lepton mass terms arise from the Yukawa coupling terms of the superpotential
Eq.~(\ref{9.4}).

If $m_0$ is large, the squarks and sleptons are nearly degenerate, and even if
$m_0 = 0$ (as in the No-scale models), the squarks are still approximately
degenerate since the $f_3$ term dominates.  This is indeed what is assumed in
the MSSM.  If $m_0$ is small, however, one would not expect the sleptons to be
degenerate.

One of the remarkable features of Eqs.~(\ref{12.2a},\ref{12.2c}) is that

\begin{equation}
m_{{\tilde c}}^2 - m_{{\tilde u}}^2 = m_c^2 - m_u^2 \label{12.4}
\end{equation} 

\begin{figure}[htb]
\vspace{2.0in}
\caption{Contributions to $K_L \rightarrow \mu^+ \mu^-$.  The left diagram (the
usual GIM term) is of size $(m_c^2 - m_u^2)/M_W^2$ while the right diagram (the
additional SUSY term) is of size $(m_{{\tilde c}}^2 - m_{{\tilde
u}}^2)/M_{{\tilde W}}^2$~or~$(m_{{\tilde c}}^2 - m_{{\tilde u}}^2)
(m_{{\tilde q}}^2)$.}
\label{6}
\end{figure}

\noindent
Condition (\ref{12.4}) guarentees a super GIM mechanism for suppressing flavor
changing neutral interactions.  Fig.~\ref{6} shows this phenomena for the case
of
$K_L \rightarrow \mu^+ \mu^-$ (first discussed in global SUSY models by
Dimopoulos and Georgi \cite{29}).  Were it not for Eq.~(\ref{12.4}), the Wino
contribution to $K_L \rightarrow \mu^+ \mu^-$ would be very large since
$m_{{\tilde q}} \r 100$~GeV.
Thus there is an enormous cancelation in the squark mass differences.  This
cancelation can be
traced to the fact that at the Gut scale, each squark gets a universal
(mass)$^2$ $m_0^2$ (which may
be very large but cancels in the difference) and a mass from gauge
interactions (which govern the $m_{1/2}^2$~and~$D$ term contributions) that
are generation blind.  The universal nature of $m_0^2$ arises from conditions
(i) and (iv) of Sec.~9 which guarentees that the super Higgs fields
communicate with the physical fields in a generation independent fashion.

\vglue 0.2cm
{\elevenit\noindent (iii) Squarks $(i = 3)$ and Radiative Breaking Masses}
\vglue 0.1cm

The ${\tilde b}_R$ squark behaves as the first two generation $d$-squarks with
mass given by Eq.~(\ref{12.2d}).  However, as can be seen from
Eqs.~(\ref{11.22}) and
(\ref{11.23}), the ${\tilde b}_L$, ${\tilde t}_L$~and~${\tilde t}_R$ are
strongly
effected by the large $t$-quark Yukawa coupling.  One finds for ${\tilde b}_L$
the result

\begin{eqnarray}
m_{{\tilde b}L}^2 = \frac{1}{2}~m_0^2 + m_b^2 &+& \frac{1}{2}~m_{{\tilde U}}^2
+ {\tilde \alpha}_G [(4/3) f_3 + (1/15) f_1] m_{1/2}^2 \nonumber \\
&+& (-\frac{1}{2}~+ \frac{1}{3}~\sin^2 \theta_W) M_Z^2 \cos
2\beta \label{12.5}
\end{eqnarray} 

\noindent
Further, the $t$-quark Yukawa coupling contribution to the $A$ term of
Eq.~(\ref{9.2})
and the $\mu H_1 H_2$ contribution to Eq.~(\ref{9.2}) cause mixing between the
${\tilde
t}_L$~and~${\tilde t}_R$ states in the $t$-squark mass matrix:

\begin{equation}
\left( \begin{array}{cc} m_{{\tilde t}_L}^2 & m_t (A_t + \mu~ctn~\beta) \\
m_t (A_t + \mu~ctn~\beta) & m_{{\tilde t}_R}^2
\end{array}\right) \label{12.6}
\end{equation} 

\noindent
where $m_t (Q) = \lambda_t (Q) v_2$~and~$A_t$ is the $t$-quark $A$
parameter at the electroweak scale.  The solution of the RGE yields

\begin{equation}
m_{{\tilde t}_L}^2 = m_Q^2 + m_t^2 + [(-1/2) + (2/3) \sin^2 \theta_W] M_Z^2
\cos 2\beta \label{12.7a}
\end{equation} 

\begin{equation}
m_{{\tilde t}_R}^2 = m_U^2 + m_t^2 + (-2/3) \sin^2 \theta_W M_Z^2 \cos 2
\beta \label{12.7b}
\end{equation} 

\noindent
where

\begin{eqnarray}
m_U^2 = \frac{1}{3}~m_0^2 &+& \frac{2}{3}~f A_0 m_{1/2} - \frac{2}{3}~k A_0^2
+ \frac{2}{5}~h m_0^2 \nonumber \\
&+& [\frac{2}{3}~e + {\tilde \alpha}_G (\frac{8}{3}~f_3 - f_2 +
\frac{1}{3}~f_1)] m_{1/2}^2 \label{12.7c}
\end{eqnarray} 

\begin{eqnarray}
m_Q^2 = \frac{2}{3}~m_0^2 &+& \frac{1}{3}~f A_0 m_{1/2} - \frac{1}{3}~k A_0^2
+ \frac{1}{3}~h m_0^2 \nonumber \\
&+& [\frac{1}{3}~e + {\tilde \alpha}_G (\frac{8}{3}~f_3 + f_2 -
\frac{1}{15}~f_1)] m_{1/2}^2 \label{12.7d}
\end{eqnarray} 

\noindent
where the functions $e (t), f (t), h (t), k (t)$ depend on $t = ln
[M_G^2/Q^2]$ and the $t$-quark Yukawa coupling, but are independent of $m_0$,
$m_{1/2}$, $A_0$, $B_0$~and~$\mu_0$.  They are defined in Iba\~nez et al
\cite{33}.  We also note that the three masses of Eq.~(\ref{11.2}) that enter
into the
radiative breaking equations (\ref{11.6}) are obtained from Eqs.~(\ref{11.11}),
(\ref{11.12}), (\ref{11.15}) and (\ref{11.16}).  One finds

\begin{equation}
m_1^2 (t) = m_0^2 + \mu^2 (t) + {\rm g} m_{1/2}^2 \label{12.8a}
\end{equation} 

\begin{equation}
m_2^2 (t) = \mu^2 (t) + e (t) m_{1/2}^2 + A_0 m_{1/2} f + m_0^2 h - A_0^2
k \label{12.8b}
\end{equation} 

\begin{equation}
m_3^2 = - B_0 \mu^2 (t) + r \mu_0 m_{1/2} + s A_0 \mu_0
\label{12.8c}
\end{equation} 

\noindent
where $\mu^2 (t) = \mu_0^2 q$~and~${\rm g} (t), q (t), r (t), s (t)$ are
defined in Iba\~nez et al \cite {33}.

\vglue 0.2cm
{\elevenit\noindent (iv) Charginos and Neutralinos}
\vglue 0.1cm

As discussed in Sec.~4, supersymmetric dynamics causes $SU(2) \times U(1)$
gaugino-Higgsino mixing after $SU(2) \times U(1)$ breaking.  The mass diagonal
states are two charged Winos (charginos) and 4 neutral Majorana Zinos
(neutralinos).  The chargino masses are \cite{23}

\begin{equation}
m_{{\tilde W}_{1,2}} = \frac{1}{2}~|[4 \nu_+^2 + (\mu - {\tilde
m}_2)^2]^{\frac{1}{2}} \mp [4\nu_-^2 + (\mu + {\tilde
m}_2)^2]^{\frac{1}{2}}| \label{12.9}
\end{equation} 

\noindent
where $\sqrt{2} \nu_{\pm} = M_W (\sin \beta \pm \cos \beta)$.  The neutralino
masses are the four roots of the secular equation $f (\lambda) = 0$ \cite {23}

\begin{eqnarray}
f (\lambda) &\equiv& \lambda^4 - ({\tilde m}_1 + {\tilde m}_2) \lambda^3 +
(M_Z^2 + \mu^2 - {\tilde m}_1 {\tilde m}_2) \lambda^2 \nonumber \\
&+& [({\tilde m}_{\gamma} - \mu \sin 2\beta) M_Z^2 + ({\tilde m}_1 + {\tilde
m}_2) \mu^2] \lambda \nonumber \\
&+& [\mu {\tilde m}_{\gamma} M_Z^2 \sin 2\beta - {\tilde m}_1 {\tilde m}_2
\mu^2] = 0 \label{12.10}
\end{eqnarray} 

\noindent
where ${\tilde m}_{\gamma} = {\tilde m}_1 \cos^2 \theta_W + {\tilde m}_2 \sin^2
\theta_W$.  We label the Zino states by ${\tilde Z}_i, i = 1 \cdots 4$, where
$m_{{\tilde Z}_i} < m_{{\tilde Z}_j}$~for~$i < j$.

\vglue 0.2cm
{\elevenit\noindent (v) Higgs Bosons}
\vglue 0.1cm

Since there are two Higgs doublets, one is left with three neutral Higgs
bosons [$h$~and~$H$ (CP-even states) and $A$ (CP-odd state)] and one charged
boson $H^{\pm}$.  The tree level masses can be obtained from the mass matrix
calculated from the Higgs potential of Eq.~(\ref{11.1a}).  One finds

\begin{equation}
m_A^2 = m_1^2 + m_2^2 = 2 m_3^2/\sin 2\beta \label{12.11}
\end{equation} 

\begin{equation}
m_{H^{\pm}}^2 = m_A^2 + M_W^2 \label{12.12}
\end{equation} 

\noindent
where the last equality of Eq.~(\ref{12.11}) follows from Eq.~(\ref{11.6}).
For the
CP-even neutral Higgs, loop corrections to the light Higgs ($h$ boson) are
important since $h$ may be very light and a heavy top quark enhances the loop
corrections.  Keeping only the top sector in the loops, $m_h^2$~and~$m_H^2$
become \cite{39}:

\begin{equation}
m_{h,H}^2 = \frac{1}{2}~[M_Z^2 + m_A^2 + \varepsilon \mp \{(M_Z^2 + m_A^2 +
\varepsilon)^2 - 4 m_A^2 M_Z^2 \cos 2\beta + \varepsilon_1\}^{1/2}]
\label{12.13}
\end{equation} 

\noindent
where $\varepsilon$~and~$\varepsilon_1$ are the loop corrections.  They can be
written in terms of two $2 \times 2$ matrices $\nu_{ij}$~and~$\Delta_{ij}$
according to

\begin{equation}
\varepsilon = Tr \Delta;~\varepsilon_1 = -4 (Tr \nu \Delta + \det
\Delta) \label{12.14a}
\end{equation} 

\noindent
where

\begin{eqnarray}
\nu_{11} &=& s^2 M_Z^2 + c^2 m_A^2;~\nu_{22} = c^2 M_Z^2 + s^2 m_A^2;~\nu_{12}
=
\nu_{21} = sc (M_Z^2 + m_A^2) \nonumber \\
\Delta_{11} &=& x \mu^2 y^2 z;~\Delta_{12} = x \mu y (w + A_t yz) =
\Delta_{21} \nonumber \\
\Delta_{22} &=& x (v + 2 A_t y w + A_t^2 y^2 z) \label{12.14b}
\end{eqnarray} 

\noindent
and

\begin{eqnarray}
x &=& \frac{3 \alpha_2}{4\pi}~\frac{m_t^4}{M_W^2 s^2}~;~y = \frac{A_t +
\mu~ctn~\beta}{m_{{\tilde
t}_1}^2 - m_{{\tilde t}_2}^2}~;~z = 2 - w \frac{m_{{\tilde t}_1}^2 + m_{{\tilde
t}_2}^2}{m_{{\tilde
t}_1}^2 - m_{{\tilde t}_2}^2} \nonumber \\
w &=& ln (m_{{\tilde t}_1}^2/m_{{\tilde t}_2}^2);~v = ln (m_{{\tilde t}_1}^2
m_{{\tilde t}_2}^2/m_t^4) \label{12.14c}
\end{eqnarray} 

\noindent
and $(s, c) = (\sin \beta, \cos \beta)$.  Large correction can arise for the
light Higgs $h$ boson for large $m_t$ due to the $m_t^4$ factor in the $x$
parameter.  Loop corrections also exist for the $A$~and~$H^{\pm}$ boson but
these are small effects unless $A$~and~$H^{\pm}$ are very light.

\vglue 0.5cm
{\elevenbf \noindent 13.  MSSM as a Low Energy Approximation}
\vglue 0.4cm

In the previous section, expressions were given allowing one to calculate each
of the
SUSY masses in terms of the Gut scale parameters.  In addition, the mass
parameters
$m_1^2$, $m_2^2$, $m_3^2$ entering in the radiative breaking equations
(\ref{11.8})
were similarly expressed allowing one to use these relations to eliminate
$B_0$~and~$\mu_0$.  (Since the masses are also expressed in terms of the VEVs
$v_1$~and~$v_2$, the loop corrections of Eq.~(\ref{11.1b}) can also be
calculated.)
Thus the 32 SUSY masses can be expressed in terms of four parameters $m_0$,
$m_{1/2}$
(or alternately the gluino mass, $m_{{\tilde g}} = (\alpha_3/\alpha_G) m_{1/2}
\simeq
3.0~m_{1/2}$), $A_0$ (or alternately $A_t$, the $t$-quark $A$ parameter at the
electroweak scale) and $\tan \beta$.

It is interesting to compare these results of supergravity grand unification
with the
MSSM.  As discussed in Sec.~4, the MSSM assumes Eq.~(\ref{12.1}) for the
gaugino
masses as well as degeneracy of the first two generations of squarks [which was
seen
to be a good approximation to Eqs.~(\ref{12.2a})-(\ref{12.2d})].  However,
there are some places
where the MSSM is not a good approximation to supergravity models:

\begin{description}
\item (i) The third generation of squarks are badly split and except for
${\tilde
b}_R$ not degenerate with the first two generations.  (The usual MSSM analysis
of data
assume all six squarks (or sometimes five squarks) are degenerate.)

\item (ii) The sleptons will not be degenerate if $m_0$ is small (which is the
case
for the No-scale model).  In addition, the MSSM depends on more parameters (and
hence
is less predictive) than the supergravity models.  For example,

\item (iii) The chargino and neutralino masses depend on $\mu$, which is
arbitrary in
the MSSM but is determined (by radiative breaking) in supergravity models.

\item (iv) The $A$ Higgs boson mass, which by Eqs.~(\ref{12.12}) and
(\ref{12.13})
determine the other three Higgs boson masses, is arbitrary in the MSSM but
depends on
the same four parameters in supergravity models [see Eqs.~(\ref{12.11}) and
(\ref{12.8c})] and hence is correlated with other masses.
\end{description}

\begin{figure}[htb]
\vspace{4.5in}
\caption{Predictions of 32 SUSY masses for supergravity grand unification for
the case
$m_t = 150$~GeV, $m_0 = 600$~GeV, $m_{{\tilde g}} = 160$~GeV, $\tan\beta =
1.73$, $A_t
= 0.0~(\mu < 0)$.  These parameters uniquely specify all masses.  The first
column is
the sleptons and first two generations of squarks, the second column is the
third
generation of squarks, the third column is the charginos and neutralinos, and
the last
column the Higgs bosons.}
\label{7}
\end{figure}

These differences are illustrated in Fig.~\ref{7}.  One sees that the third
generation of squarks are
badly split with the lightest stop, ${\tilde t}_1$, being the lightest squark.
Also there are three
light Winos and Zinos and three heavy ones, and one very light Higgs ($h$) and
three heavy ones ($H,
A, H^{\pm}$).  The sleptons are nearly degenerate with the first two
generations of
squarks since $m_0$ is fairly large for this case.

\vglue 0.5cm
{\elevenbf \noindent 14.  Threshold Corrections to Grand Unification}
\vglue 0.4cm

In the discussion of the unification of the coupling constants in Sec.~6, it
was
assumed that all SUSY particles had a common mass $M_S$.  As we have seen, the
SUSY
spectrum is split ranging from below $M_Z$ up to about 1 TeV.  As a consequence
there
are threshold corrections to the RGE that describe the unification of the
couplings.
Similarly at the Gut scale, as seen in the $SU(5)$ example of Sec.~10, we may
expect
an array of superheavy non-degenerate states with masses of $O (M_G)$ which
will also
have threshold corrections.  These corrections can modify the grand unification
results and some analyses of these problems have been given
\cite{40}-\cite{43}.

We consider first the case of the low energy thresholds at $M_S$.  For a fixed
Gut
model defined by specific Gut parameters $m_0$, $m_{1/2}$ etc., we saw we can
calculate all the SUSY masses and thus be able to insert the thresholds in the
RGE
running of the coupling constants.  Each threshold causes a kink in
$\alpha_i^{-1}$
(Fig.~\ref{8}) which shifts the positions of $M_G$~and~$\alpha_G$.  However,
the positions of each
SUSY mass depends on $M_G$~and~$\alpha_G$ (from the running of the RGE) as can
be seen in Sec.~12.
Hence one has an additional dependence of the final grand unification point on
the position of
the threshold.  Thus a complicated numerical analysis is needed to deal with
this.

It is possible to parameterize the SUSY thresholds phenomenologically
\cite{42}.  Thus integrating
Eq.~(\ref{6.4}) from $\mu = M_G$~to~$\mu = M_Z$ gives:

\begin{equation}
\alpha_i^{-1} (M_Z) = \alpha_G^{-1} + b_i t_0 - \Delta_i + 2-{\rm
loop} \label{13.1}
\end{equation} 

\noindent
where $t_0 = ln [M_G/M_z]/2\pi$~and~$\Delta_i$ are the SUSY threshold pieces,

\begin{equation}
\Delta_i = \frac{1}{2\pi} \sum b_i^{(a)} ln (M_a/M_z) \label{13.2}
\end{equation} 

\noindent
Here $b_i^{(a)}$ is the contribution of the $a$th particle to the beta
functions
$\beta_i = b_i \alpha_i^2/2\pi$ (each particle decoupling at its mass as one
proceeds
downwards from $M_G$).  The $\Delta_i$ are of course functions of the Gut
parameters
and can be explicitly calculated using Sec.~12.  Instead, one may use a
phenomenological parameterization defined by three masses $M_i$, $i = 1, 2, 3$:

\begin{equation}
\sum b_i^{(a)} ln (M_a/M_Z) \equiv (b_i^{MSSM} - b_i^{SM}) ln
(M_i/M_Z) \label{13.3}
\end{equation} 

\noindent
Clearly, if all SUSY particles are degenerate, $M_i = M_S$, but otherwise the
$M_i$
are distinct.  Now one may use the place where $\alpha_1$~and~$\alpha_2$
intersect to
define the grand unification point $\mu = M_G$ and the value of $\alpha_G$.
Then by
requiring also $\alpha_3 (M_G) = \alpha_G$ one can predict the value of
$\alpha_3
(M_Z)$ (which can then be compared with experiment).  The threshold
contributions to
$\alpha_3 (M_z)$ are

\begin{equation}
25~ln (M_1/M_Z) - 100~ln (M_2/M_Z) + 56~ln (M_3/M_Z) \equiv - 19~ln
(M_S^{eff}/M_Z) \label{13.4}
\end{equation} 

\noindent
which defines an effective $M_S$.  Note however, that $M_S^{eff}$ is different
from
the average SUSY mass since it is an average of logarithms weighted by $\beta$
function factors, one of which is negative.  Thus it is even possible to have
$M_S^{eff} < M_Z$.  In general one finds that these threshold corrections are
about
as large as the current error flags, and hence will become more important as
experimental errors decrease.

\begin{figure}[htb]
\vspace{3.5in}
\caption{Schematic diagram showing effects of SUSY mass thresholds.  The values
of
$M_G$~and~$\alpha_G$ are shifted when threshold effects are taken into
account.}
\label{8}
\end{figure}

An important point to remember in considering the above analysis is that the
energy
scale of supersymmetry breaking is the same of $SU(2) \times U(1)$ breaking.
This
can be seen, for example, in Fig.~\ref{7}, where the mass spectrum of the third
generation
squarks, the Winos and Zinos etc. are not $SU(2) \times U(1)$ invariant.  (In
fact,
the ${\tilde Z}_{1,2}$~and~${\tilde W}_1$ may lie below $M_Z$.)  Thus in
proceeding
up in energy from $M_Z$, one cannot assume $SU(2) \times U(1)$ invariance holds
in
the SUSY particle region, as is sometimes done.

We turn next to consider the Gut scale thresholds.  The physics at the Gut
scale is
unknown, and so treatment of threshold effects here are more speculative.  In
order
to learn the general nature of these effects we consider the $SU(5)$ model of
Eqs.~(\ref{10.4}) and (\ref{10.5}).  From explicit calculation of the mass
matrix one
finds that the superheavy particles arising from the breaking of $SU(5)$ are
the
following:

\begin{description}
\item (i) Two color Higgs triplet chiral multiplets transforming under $SU(3)_C
\times SU(2)_L$ as (3,1) and ($\overline{3}$,1) with mass $M_{H_3} = 5
\lambda_2 M$.

\item (ii) Two massive vector multiplets (consisting each of a massive vector
boson,
Dirac spinor and hermitian scalar) transforming as (3,2) and ($\overline{3}$,2)
with
mass $M_V = 5 \sqrt{2} g M$, where $\alpha_G = g^2/4\pi$.

\item (iii) Those components of the chiral multiplets of $\sum_Y^X$ not
absorbed
by the vector bosons in their mass growth.  These are two degenerate
representations
transforming as (8,1) and (1,3) with masses $M_{\Sigma}^8 = 5 \lambda_1 M/2 =
M_{\Sigma}^3$ and a SM singlet with mass $M_{\Sigma}^0 = \lambda_1 M/2$.
\end{description}

Upper bounds on the $\lambda_{1,2}$ arise by requiring that the model stay
within
the perturbative domain.  We take this to mean $\lambda_{1,2} \leq 2$, or
$\alpha_{\lambda_{1,2}} = \lambda_{1,2}^2/4\pi \l 1/3$.  Then since
$\alpha_G^{-1}
\simeq 25$ one has

\begin{equation}
\frac{M_{H_3}}{M_V}~= \frac{\lambda_2}{\sqrt{2}g}~\l~2; \quad
\frac{M_{\Sigma}^{8,3}}{M_V}~\l~1; \quad \frac{M_{\Sigma}^0}{M_V}~\l~0.2
\label{13.5}
\end{equation} 

\noindent
Thus the superheavy particles cannot get much larger than $M_V$.  Since the
superheavy
particles are expected to be $O (M_G)$ we will also require $\lambda_{1,2} >
0.01$~or~$\alpha_{\lambda_{1,2}}~\r~10^{-5}$.

\begin{figure}[htb]
\vspace{3.25in}
\caption{Grand unification including Gut thresholds of the model of Eq.~(77).
The allowed region
is that bounded by the solid quadrilateral.}
\label{9}
\end{figure}

One may now run the RGE up to the Gut scale, taking into account the additional
Gut
thresholds.  An accurate analysis requires inclusion of two loop effects.
Fig.~\ref{9}
\cite{43} shows the effects of the Gut scale thresholds neglecting the low
energy SUSY
thresholds.  Grand unification implies a correlation between $M_{H_3}$~and~
$\alpha_3 (M_Z)$.
For $\lambda_1 > 0.01$~or~$\lambda_2 < 2$, the horizontal line moves downwards.
The effective
$M_S$ is assumed to obey the bounds 30 GeV $< M_S < 1$ TeV.  The solid
quadrilateral thus
represents the allowed region of grand unification.  The model thus predicts
$\alpha_3 (M_z) <
0.135$ consistent with the experimental value of $\alpha_3 (M_Z) = 0.118 \pm
0.007$.  At the
1$\sigma$ upper bound of $\alpha_3 (M_Z) = 0.125$, one finds $M_{H_3} < 2
\times 10^{17}$~GeV.
Clearly, an accurate determination of $\alpha_3 (M_Z)$ would allow further
tests of this model as
$M_{H_3}$ is related to the proton decay rate.

\vglue 0.5cm
{\elevenbf \noindent 15.  Models with Proton Decay}
\vglue 0.4cm

The analysis to obtain predictions of supergravity Gut models can proceed as
follows:
One calculates the 32 SUSY masses in terms of the $4 + 1$ parameters $m_0$,
$m_{1/2}$,
$A_0$, $\tan \beta$~and~$m_t$.  One then varies these parameters over their
entire range
subject to the conditions that (i) Radiative breaking of $SU(2)_L \times
U(1)_Y$ occurs,
and (ii) Experimental bounds on SUSY masses are not violated.  One then gets
the allowed
mass bands and correlations between SUSY masses.  These are the experimental
predictions
of the model.

It is possible to reduce the size of the allowed parameter space by imposing
additional
constraints.  Three possibilities that have been considered are:  (i)
Constraints due to
experimental bounds on proton decay, (ii) Cosmological constraints from SUSY
candidates
for dark matter, and (iii) Constraints from conditions on Yukawa couplings at
the Gut
scale [e.g. $\lambda_b (M_G) = \lambda_{\tau} M_G)$].  We consider the first of
these
here.

There are two main decay modes possible in SUSY Gut models:  $p \rightarrow e^+
+
\pi^0$~and~$p \rightarrow \overline{\nu} + K^+$.  The current 90\% CL bounds
from
Kamiokande and IMB are \cite{44}

\begin{equation}
\tau (p \rightarrow e^+ \pi^0) > 5.5 \times 10^{32} yr \label{14.1a}
\end{equation} 

\begin{equation}
\tau (p \rightarrow \overline{\nu} K) > 1.0 \times 10^{32} yr \label{14.1b}
\end{equation} 

\noindent
The decay $p \rightarrow e^+ \pi^0$ can occur in both SUSY and non-SUSY models.
It
proceeds through the superheavy vector bosons ($X^{\mu}$, $Y^{\mu}$ in $SU(5)$
theory)
with mass $M_V$ as shown in Fig.~\ref{10}.  For SUSY models, the lifetime for
$p \rightarrow e^+
\pi^0$ is given by \cite{42}:

\begin{figure}[htb]
\vspace{2.25in}
\caption{Proton decay diagram for the mode $p \rightarrow e^+ + \pi^0$.}
\label{10}
\end{figure}

\begin{equation}
\tau (p \rightarrow e^+ \pi^0) \simeq (\frac{M_V}{3.5 \times 10^{14}~GeV})^4
10^{31 \pm
1} yr \label{14.2}
\end{equation} 

\noindent
Super Kamiokande plans to be sensitive to this mode up to about $1 \times
10^{34}$ yr.
To be observable at this level requires (at about 90\% CL) that $M_V~\l~5
\times
10^{15}$~GeV.  Since the mean $M_G$ value is $M_G \simeq 1.5 \times
10^{16}$~GeV, it is
doubtful that this model will be observable at Super Kamiokande.

The decay $p \rightarrow \overline{\nu} + K^+$ is specifically a supersymmetric
mode and
hence observation of this decay would be a real indication of the validity of
supergravity Guts.  However, it does not necessarily arise in all supergravity
models.
We consider here ``$SU(5)$-type'' proton decay models defined as follows:

\begin{description}
\item (i) The Gut group $G$ contains an $SU(5)$ subgroup [or is $SU(5)$].

\item (ii) The matter that remains light after $G$ breaks to $SU(3)_C \times
SU(2)_L
\times U(1)_Y$ at $M_G$ is embedded in the usual way in the 10 and
$\overline{5}$
representations of the $SU(5)$ subgroup.

\item (iii) After $G$ breaks, there are only two light Higgs doublets which
interact
with matter, and these are embedded in the 5 and $\overline{5}$ of the $SU(5)$
subgroup.

\item (iv) There is no discrete symmetry or fine tuning condition that forbids
the
proton decay amplitude.
\end{description}

\noindent
The above conditions can hold for a variety of groups besides $SU(5)$ e.g. $O
(10),
E_6$ etc.  Remarkably, models obeying (i) - (iv) give rise to a universal decay
amplitude \cite{45,46} for $p \rightarrow \overline{\nu} + K^+$ arising from
the exchange of
the superheavy Higgsino color triplet of mass $M_{H_3}$.  A characteristic
diagram is
shown in Fig.~\ref{11}.

\begin{figure}[htb]
\vspace{2.0in}
\caption{One diagram contributing to the decay amplitude for $p \rightarrow
\overline{\nu} + K^+$.  The Wino ``dressing'' converts quarks to squarks and
the
${\tilde H}_3$ vertices violate baryon and lepton number.  There are additional
diagrams
with $\overline{\nu}_e$, $\overline{\nu}_{\tau}$ final states.  Also the CKM
matrix
elements appear at the ${\tilde W}$ vertices allowing all three generations to
enter in
the loop.}
\label{11}
\end{figure}

Proton decay is characteristic of grand unified models and one might ask under
what
circumstances it can be suppressed.  It appears difficult to find a discrete
symmetry
that can suppress the $p \rightarrow \overline{\nu} K^+$ decay without
introducing
additional Higgs doublets (which would ruin grand unification) or produce some
other
illness.  The only other natural way of suppressing the decay is by violating
condition
(ii).  This is done in the flipped $SU(5) \times U(1)$ model where the
interchanges $u
\leftrightarrow d$~and~$e \leftrightarrow \nu$ in the particle embeddings
suppresses proton
decay.  We will come back to this model later.

Conditions (i) - (iv) imply the existance of an $SU(5)$ invariant contribution
to the
Yukawa part of the superpotential of

\begin{equation}
W_Y = \lambda_{ij}^1 \varepsilon_{XYZWU} H_1^X M_i^{YZ} M_j^{WU} +
\lambda_{ij}^2
\overline{H}_{2X} \overline{M}_{iY} M_j^{XY} \label{14.3}
\end{equation} 

\noindent
where we have used the notation of Sec.~10.  It is the superheavy
$H_1^a$~and~$\overline{H}_{2a}$ couplings in Eq.~(\ref{14.3}) which give rise
to the
$p$-decay interaction.  One may eliminate these fields to obtain an effective
dimension
five interaction scaled by $1/M_{H_3}$.  The total decay rate is

\begin{equation}
\Gamma (p \rightarrow \overline{\nu} K^+) = \sum_i \Gamma (p \rightarrow
\overline{\nu}_i K^+) \label{14.4}
\end{equation} 

\noindent
where $\overline{\nu}_i = (\overline{\nu}_e, \overline{\nu}_{\mu},
\overline{\nu}_{\tau}$).  The calculation of $\Gamma (p \rightarrow
\overline{\nu} K^+)$
is rather lengthy.  We state here the final result \cite{46} for reference.
The first
generation contributions are negligible, and so one may limit $i$ to $i = \mu,
\tau$.
One can write

\begin{equation}
\Gamma (p \rightarrow \overline{\nu}_i K^+) = C (\frac{\beta_p}{M_{H_3}})^2
|A|^2
|B_i| \label{14.5}
\end{equation} 

\noindent
where $C$ is a chiral current algebra factor (Chadha and Daniels \cite{45})

\[
C = \frac{m_N}{32 \pi f_{\pi}^2}~[(1 + \frac{m_N (D + F)}{m_B})~(1 -
\frac{m_K^2}{m_N^2})]^2
\]

\noindent
($D = 0.76$, $F = 0.48$, $f_{\pi} = 139$~MeV, $m_N = 938$~MeV, $m_B =
1154$~MeV, $m_K =
495$~MeV).  The factor $A$ is

\begin{equation}
A = \frac{\alpha_2^2}{2 M_W^2}~m_s m_c V_{21}^{\dagger} V_{21} A_L A_S
\label{14.6}
\end{equation} 

\noindent
where $V_{ij}$ are CKM elements, $A_{L,S}$ are renormalization group factors to
take the
operator at the Gut scale to the proton scale (Ellis et al \cite{45}).  We find
$A_L =
0.283$, $A_S = 0.833$.  The quantity $\beta_p$ describes the quark content of
the proton,

\begin{equation}
\beta_p U_L^{\gamma} = \varepsilon_{abc} \varepsilon_{\alpha\beta} <o |
d_{aL}^{\alpha}
u_{bL}^{\beta} u_{cL}^{\gamma} | p> \label{14.7}
\end{equation} 

\noindent
where $U_L^{\gamma}$ is the proton wave function, $d_{aL}^{\alpha}$,
$u_{bL}^{\beta}$ are quark
operators ($\alpha,\beta =$~spinor indices).  $\beta_p$ has been evaluated by
lattice
gauge theory \cite{47}:

\begin{equation}
\beta_p = (5.6 \pm 0.5) \times 10^{-3}~{\rm GeV}^3 \label{14.8}
\end{equation} 

\noindent
The loop integral $B_i$ may be written as

\begin{equation}
B_i = \frac{m_i^d V_{i1}^{\dagger}}{m_s V_{21}^{\dagger}}~[P_2 B_{2i} +
\frac{m_t V_{31} V_{32}}{m_c
V_{21} V_{22}}~P_3 B_{3i}]~\frac{1}{\sin 2\beta} \label{14.9}
\end{equation} 

\noindent
where $m_i^d$ are $d$-quark masses and $B_{ji}$, the contribution of $j$th
generation
particles in the loop to the amplitude for $p \rightarrow \overline{\nu}_i K^+$
may be
written as

\begin{equation}
B_{ji} = F ({\tilde u}_i, {\tilde d}_j, {\tilde W}) + ({\tilde d}_j \rightarrow
{\tilde
e}_j) \label{14.10}
\end{equation} 

\noindent
where

\begin{eqnarray}
&&F ({\tilde u}_i, {\tilde d}_j, {\tilde W}) = [E \cos \gamma_- \sin \gamma_+
{\tilde f}
({\tilde u}_i, {\tilde d}_j, {\tilde W}_1) + \nonumber \\
&&\cos \gamma_+ \sin \gamma_- {\tilde f} ({\tilde u}_i, {\tilde d}_j, {\tilde
W}_2)]~-
\frac{1}{2}~\frac{\delta_{i3} m_i^u \sin 2 \delta_{ui}}{\sqrt{2}~M_W \sin
\beta}~[\{E \sin
\gamma_- \sin \gamma_+ f ({\tilde u}_{i1}, {\tilde d}_j, {\tilde W}_1)
\nonumber \\
&&\qquad - \cos \gamma_- \cos \gamma_+ f ({\tilde u}_{i1}, {\tilde d}_j,
{\tilde W}_2)\}
- \{{\tilde u}_{i1} \rightarrow {\tilde u}_{i2}\}] \label{14.11}
\end{eqnarray} 

\noindent
with the definitions $\gamma_{\pm} = \beta_+ \pm \beta_-$,

\begin{equation}
\sin 2 \beta_{\pm} = \frac{(\mu \mp {\tilde m}_2)}{[4\nu_{\pm}^2 + (\mu \pm
{\tilde
m}_2)^2]^{1/2}};~\sqrt{2}~\nu_{\pm} = M_W (\sin \beta \pm \cos \beta)
\label{14.12}
\end{equation} 

\begin{equation}
\sin 2 \delta_{u3} = \frac{-2 (A_t + \mu~ctn~\beta) m_t}{m_{{\tilde t}_i}^2 -
m_{{\tilde
2}_2}^2} \label{14.13}
\end{equation} 

\begin{equation}
E = \left\{ \begin{array}{ll} +1, & \sin 2 \beta > \mu {\tilde m}_2/M_W^2 \\
-1, & \sin
2\beta < \mu {\tilde m}_2/M_W^2\end{array} \right. \label{14.14}
\end{equation} 

\noindent
and

\begin{eqnarray}
{\tilde f} ({\tilde u}_i, {\tilde d}_j, {\tilde W}_k) &=& \sin^2 \delta_{ui} f
({\tilde
u}_{i1}, {\tilde d}_j, {\tilde W}_k) \nonumber \\
&+& \cos^2 \delta_{ui} f ({\tilde u}_{i2}, {\tilde d}_j, {\tilde W}_k)
\label{14.15}
\end{eqnarray} 

\noindent
with ${\tilde u}_{i1}$~and~${\tilde u}_{i2}$ being the $u$-squark mass states
and

\begin{equation}
f (a, b, c) = \frac{m_c}{m_b^2 - m_c^2}~[\frac{m_b^2}{m_a^2 - m_b^2}~ln
(\frac{m_a^2}{m_b^2})~- (m_b \rightarrow m_c)] \label{14.16}
\end{equation} 

\noindent
Again $j = 1$ gives a neglegible contribution and we have include L-R mixing
only for the
$t$-squarks.

In Eq.~(\ref{14.9}), the phases $P_j = e^{i \alpha_j}, j = 2,3$ appear.  These
are new
CP violating phases (independent of the CKM phase) that enter in the dimension
five
operators.  Thus there are two limiting cases:  $P_3/P_2 = - 1$, destructive
interference between the second and third generation contributions to the loop,
and
$P_3/P_2 = + 1$, constructive interference.

It is convenient to define the quantity

\begin{equation}
B \equiv [|B_2|^2 + |B_3|^2]^{\frac{1}{2}}~[M_S/10^{2.4}~GeV]^{0.33} \times
10^6~GeV^{-1} \label{14.17}
\end{equation} 

\noindent
and using Eq.~(\ref{6.8}) (to account for the anti-correlation between
$M_S$~and~$M_G$),
the Kamiokande experiment bound, Eq.~(\ref{14.1b}), can be written as

\begin{equation}
B~\l~100~(\frac{M_{H_3}}{M_G})~{\rm GeV}^{-1} \label{14.18}
\end{equation} 

\noindent
for $\beta_p = 5.6 \times 10^{-3}$~GeV$^3$.  The quantity $B$ is a function of
the SUSY
masses (squarks, sleptons, Winos) and so the experimental bound on $B$ becomes
a
constraint on the SUSY mass spectrum.

\vglue 0.5cm
{\elevenbf \noindent 16.  Predictions for Models with Proton Decay}
\vglue 0.4cm

We now discuss what predictions can be made for the SUSY mass spectrum for
models
possessing $SU(5)$-type proton decay.  We will see, in fact, that with
sensitivity
expected in future underground and accelerator experiments, it should be
possible to
test whether these models are right or wrong.

Our procedure is to allow the  basic parameters, $m_0$, $m_{1/2}$, $A_0$, $\tan
\beta$~and~$m_t$ to vary arbitrarily subject to the following constraints:  (i)
Radiative breaking of $SU(2)_L \times U(1)_Y$ occurs.  (ii) Current
experimental bounds
on SUSY masses are obeyed.  (iii) Two theoretical constraints are imposed:  (a)
No
extreme fine tuning of parameters are allowed which we quantify by requiring
$m_{{\tilde
q}}$, $m_{{\tilde g}} < 1$~TeV, and (b) $M_{H_3}$ is $O (M_G)$ which we
quantify by
requiring $M_{H_3}/M_G < 10$.

Condition (iiia) accepts fine tuning up to about the 1\% level which is about
the limit
of loop corrections.  Condition (iiib) implies $M_{H_3}~\l~2 \times
10^{17}$~GeV.  Any
larger value of $M_{H_3}$ would need to include Planck physics corrections and
the
supergravity Gut theory would not be self-contained.  Note also that in the
simple Gut
model discussed in Sec.~13, Fig.~9, the experimental value of $\alpha_3 (M_Z)$
requires
a bound of this size for grand unification to occur.

We review briefly now the procedure of calculation:  (i) Run the one loop SUSY
RGE from
$M_G$ to $M_Z$ and impose radiative breaking on the Higgs potential.  (ii)
Calculate
all 32 SUSY masses as a function of $m_0$, $m_{1/2}$, $A_0$,
$\tan\beta$~and~$m_t$.
(iii) Limit the parameter space so that the experimental bounds on SUSY masses
are
obeyed.  (iv) Limit the parameter space so that proton decay bounds are obeyed
(i.e. $B
< 1000$~GeV$^{-1}$).  (v) Allow parameters to vary over the remaining allowed
regions
and obtain the allowed bands of SUSY masses and the correlations between
masses.  These
are the predictions that can be checked experimentally.

The calculations discussed below are made under the following conditions:  (1)
In (i)
above we neglect loop corrections to the Higgs potential.  As discussed above,
these
are generally small \cite{37}.  (2) However, in (ii) above we do include the
loop
corrections to the light Higgs mass $m_h$ as these may be large, particularly
for large
$m_t$.  (3) In calculating $B$, we assume the distructive interference
possibility i.e.
$P_2/P_3 = -1$.  This minimizes as much as possible the effects of proton decay
on the
SUSY mass spectrum.  (4) Central values of CKM matrix elements are used:
$V_{31} =
0.011$, $V_{32} = - 0.042$, $V_{13} = - 0.002$.  (5) Only the $t$-quark Yukawa
coupling
was included.  This is a good approximation for $\tan \beta < 10$, which we
will see is
generally valid.

The proton decay amplitude function $B$ of Eq.~(\ref{14.17}) is a complicated
function
of SUSY masses.  We can get a qualitative picture of what the proton decay
constraint
implies by considering only the second generation contribution in the limit of
large
$m_0$ (i.e. $m_0^2 \gg M_Z^2, {\tilde W}_2^2)$.  Then $B$ takes the simple form

\begin{equation}
B_2 \simeq - \frac{2 \alpha_2}{\alpha_3 \sin 2\beta}~~\frac{m_{{\tilde
g}}}{m_{{\tilde
q}}^2}~\times 10^6 \label{15.1}
\end{equation} 

\noindent
where the RGE give $m_{{\tilde q}}^2 \cong m_0^2 + (0.65) m_{{\tilde g}}^2$.
An upper
bound on $| B_2 |$ generally implies (1) an upper bound on $m_{{\tilde g}}$
(unless
$m_{{\tilde g}}$ is very large, i.e. $m_{{\tilde g}}^2 \gg m_0^2$ as then $|
B_2 | \sim
1/m_{{\tilde g}}$), (2) a lower bound on $m_0$ so that $m_{{\tilde q}}^2$ does
not get
too small (again unless $m_{{\tilde g}}$ is very large), and (3) an upper bound
on
$\tan \beta$ so that $\sin 2\beta$ does not get too small.

\begin{figure}[htb]
\vspace{3.0in}
\caption{$B$ as a function of $m_0$ for $m_t = 125$~GeV, $m_{{\tilde g}} =
160$~GeV,
$\tan \beta = 1.73$, $\mu < 0$.  The low $m_0$ cutoff for $A_t < 0$ is due to
${\tilde
t}_1$ turning tachyonic.}
\label{12}
\end{figure}

\begin{figure}[htb]
\vspace{3.25in}
\caption{$B$ as a function of $m_{{\tilde g}}$ for $m_t = 125$~GeV, $A_t = -0.6
m_0$,
$\tan \beta = 1.73$, $\mu < 0$.  An upper bound $m_0 < 1$~TeV implies a lower
bound on
$m_{{\tilde g}}$.}
\label{13}
\end{figure}

The above qualititive features are exhibited in Fig.~\ref{12} and Fig.~\ref{13}
\cite{48}.  These
figures also exhibit the fact that current proton decay data for $p \rightarrow
\overline{\nu} K^+$
requires $M_{H_3}~\r~1 \times 10^{16}$~GeV $(M_{H_3}/M_G~\r~2/3)$.  In the Gut
model of Sec.~10,
then, Eq.~(\ref{13.5}) and the bound on $M_V$ of Eq.~(\ref{14.2}) would make it
unlikely that Super
Kamiokande would detect the $p \rightarrow e^+ \pi^0$ mode (and if it did, the
model predicts that
the $p \rightarrow \overline{\nu} K^+$ mode would be very copious).  The bound
on $M_{H_3}$ also
implies $\alpha_3 (M_Z)~\r~0.114$ in this model.

It is convenient to consider the results as a function of $M_{H_3}$:

\begin{enumerate}
\item $M_{H_3}/M_G < 3$

For this case, where $M_{H_3}$ is relatively small, one finds that
$m_0~\r~500$~GeV and
$m_{{\tilde g}}~\l~450$~GeV (as illustrated in Figs.~14 and 15).  In addition
one finds
$| A_t |/m_0~\l~1.5$~and~$1.1~\l~\tan \beta~\l~5$.  One expects therefore that
the
squarks (with the possible exception of ${\tilde t}_1$) and probably also
gluinos will
require the SSC or LHC to be detected.  The $t$ quark and $h$ boson are limited
to $m_t
< 180$~GeV and $m_h < 110$~GeV.

\item As one increases $M_{H_3}$, the lower bound on $m_0$ decreases and the
upperbound on
$m_{{\tilde g}}$ increases.  Thus for $M_{H_3}/M_G~\r~7$, $m_{{\tilde g}}$ can
saturate
its upper bound of 1 TeV (though usually a large $m_{{\tilde g}}$ requires a
small
$m_0$.)  The allowed bands on $\tan \beta$~and~$A_t$ widen a little (e.g. $\tan
\beta~\l~8$).

\item Perhaps the most remarkable results is a set of scaling relations that
hold among
the charginos, neutralinos and gluino over most of the allowed parameter space
\cite{48,49}
\end{enumerate}

\begin{equation}
2 m_{{\tilde Z}_1} \cong m_{{\tilde W}_1} \cong m_{{\tilde Z}_2} \label{15.2a}
\end{equation} 

\begin{equation}
m_{{\tilde W}_1} \simeq \frac{1}{3}~m_{{\tilde g}}~{\rm for}~\mu <
0;~m_{{\tilde W}_1}
\simeq \frac{1}{4}~m_{{\tilde g}}~{\rm for}~\mu > 0 \label{15.2b}
\end{equation} 

\begin{equation}
m_{{\tilde W}_2} \cong m_{{\tilde Z}_3} \cong m_{{\tilde Z}_4} \gg m_{{\tilde
Z}_1} \label{15.2c}
\end{equation} 

\noindent
In addition the Higgs boson masses obey

\begin{equation}
m_H^0 \cong m_A \cong m_{H^{\pm}} \gg m_h \label{15.3}
\end{equation} 

\noindent
Eqs.~(\ref{15.2a}) and (\ref{15.3}) are illustrated in the example of Fig.~7.

Eqs.~(\ref{15.2a})-(\ref{15.3}) are a consequence of the radiative breaking
equations
Eqs.~(\ref{11.6}), which determine the parameter $\mu$ \cite{49}.  From
Eqs.~(\ref{12.8a})-(\ref{12.8c}), one sees that the parameters entering there
depend on $m_0^2$,
$\mu^2$~and~$m_{{\tilde g}} \cong 3 m_{1/2}$.  Since the proton decay
constraints require that $m_0$
be relatively large (or if $m_0$ is small, $m_{{\tilde g}}$ is large), the
$\mu^2$ contribution must
cancel these large numbers if the r.h.s. of the first equation in (\ref{11.6})
is to add
up to $\frac{1}{2}~M_Z^2$.  Thus in general, one has $\mu^2 \gg M_Z^2$,
${\tilde m}_2^2$
(a result that often holds also even in the No-scale model when $m_0 = 0$,
since then
$m_{1/2}$ is often large).  Expanding Eq.~(\ref{12.9}) in this limit gives
\cite{50}

\begin{equation}
m_{{\tilde W}_1} \cong {\tilde m}_2 - \frac{M_W^2 \sin
2\beta}{\mu}~,~m_{{\tilde W}_2}
\cong \mu + \frac{M_W^2}{\mu} \label{15.4}
\end{equation} 

\noindent
where the $O (1/\mu)$ corrections are small.  Since ${\tilde m}_2 =
(\alpha_2/\alpha_3)
m_{{\tilde g}} \cong 0.285 m_{{\tilde g}}$ one sees the origin of
Eq.~(\ref{15.2b}) since
$m_{{\tilde W}_1}$ is increased (decreased) a small amount depending on whether
$\mu$ is
negative (positive).  The secular equation (\ref{12.10}) has two light roots
and two heavy
roots in this limit:

\begin{equation}
m_{{\tilde Z}_1} \cong {\tilde m}_1 - \frac{M_Z^2 \sin 2\beta \sin^2
\theta_W}{\mu}~;~m_{{\tilde Z}_2} \cong {\tilde m}_2 - \frac{M_W \sin 2
\beta}{\mu} \label{15.5}
\end{equation} 

\begin{equation}
m_{{\tilde Z}_3, {\tilde Z}_4} \cong | \mu - \frac{1}{2}~\frac{M_Z^2}{\mu}~(1
\pm \sin
2\beta | \label{15.6}
\end{equation} 

\noindent
Since ${\tilde m}_1/{\tilde m}_2 = \alpha_1/\alpha_2 \cong 0.506$,
Eqs.~(\ref{15.4}) and
(\ref{15.5}) lead to Eq.~(\ref{15.2a}) while from Eq.~(\ref{15.4}) and
(\ref{15.6}) one
deduces Eq.~(\ref{15.2c}).  Finally from Eq.~(\ref{12.8a}),(\ref{12.8b}) one
has $m_1^2 + m_2^2 =
2\mu^2 + m_0^2 + \cdots \gg M_Z^2$, and expanding the formulae for the Higgs
boson masses,
Eqs.~(\ref{12.11})-(\ref{12.13}), one deduces Eq.~(\ref{15.3}).  We note also
that
Eq.~(\ref{15.3}) implies that the $h$ boson couplings are close to those of the
SM.

The theoretical origin of the upper bound $m_t~\l~180$~GeV (in good agreement
with the
phenomenological result from data analysis \cite{2}) is due to the left-right
mixing in the
squark mass matrices.  Thus from Eq.~(\ref{12.6}), as $m_t$ increases the lower
stop
eigenvalue $m_{{\tilde t}_1}^2$ decreases (since the off diagonal elements
increase) until
$m_{{\tilde t}_1}$ is reduced below 45 GeV (it's current experimental bound)
beyond which
$m_{{\tilde t}_1}^2$ is rapidly driven tachyonic.  Thus large $m_t$ reduces the
parameter
space by requiring $A_t$~and~$ctn~\beta$ to decrease.  But proton decay puts a
lower
bound on $ctn~\beta$ (i.e. $\tan \beta~\l~8$), and so eventually the allowed
parameter
space is exhausted.  This occurs for $m_t \simeq 180$~GeV.

To illustrate the significance of the proton decay constraint, one can examine
the class
of No-scale models (where $m_0 = 0 = A_0$) which allow proton decay.  One finds
that
present data for proton decay excludes this model except in a small corner of
the
parameter space, $m_{{\tilde g}}~\r~900$~GeV, $\tan \beta~\l~2$ \cite{51}.
Thus No-scale
models that are phenomenologically acceptable must forbid proton decay.  The
one model
that does this in a natural way is the flipped $SU(5) \times U(1)$ model
\cite{32}, which
is phenomenologically acceptable and is discussed below.

\begin{figure}[htb]
\vspace{3.0in}
\caption{The maximum value of $\tau (p \rightarrow \overline{\nu} K^+)$ for
$m_t =
150$~GeV, $\mu < 0$ as all other parameters are varied.  The three curves are
for
$M_{H_3}/M_G = 3, 6, 10$ (lower, intermediate, upper curves).  ICARUS can
detect signals
below the upper horizontal line and Super Kamiokande below the lower horizontal
line.}
\label{14}
\end{figure}

\begin{figure}[htb]
\vspace{3.0in}
\caption{The maximum value of $\tau (p \rightarrow \overline{\nu} K^+)$ for
$m_t =
150$~GeV, $\mu < 0$ subject to $m_{{\tilde W}_1} > 100$~GeV.  Curves are
labeled as in
Fig.~14.}
\label{15}
\end{figure}

Future proton decay experiments will be able to give strong tests for models
with proton decay.
Super Kamiokande will be able to detect the $p \rightarrow \overline{\nu} K^+$
mode for a
lifetime up to $2 \times 10^{33}$ yr \cite{52} and ICARUS expects to be able to
reach $5
\times 10^{33}$ yr \cite{53}.  To exhibit the reach of these experiments, we
maximize the
partial lifetime $\tau (p \rightarrow \overline{\nu} K)$ at fixed $m_0$ by
varying all
the other parameters over the entire allowed parameter space.  Fig.~\ref{14}
exhibits this for
the case $m_t = 150$~GeV for three characteristic values of $M_{H_3}$
\cite{54}.  One can
see that ICARUS can test proton decay for $m_0~\l~800$~GeV, and Super
Kamiokande for
$m_0~\l~600$~GeV.  Fig.~\ref{15} shows a similar plot but with the parameters
restricted so that
$m_{{\tilde W}_1} > 100$~GeV.  Here we can see that ICARUS can detect proton
decay over the entire
allowed parameter space, $m_0 < 1$~TeV, $M_{H_3}/M_G < 10$ (and Super
Kamiokande over the region
$m_0~\l~950$~GeV, $M_{H_3}/M_G < 10$).  Thus if ICARUS does not see proton
decay, then $m_{{\tilde
W}_1} < 100$~GeV, and the ${\tilde W}_1$ should be observable at LEP200 (and
also possibly at the
Tevatron).  Thus we have the prediction that for models with proton decay
obeying the conditions (i)
- (iii) stated at the beginning of this section, either proton decay is
observable at future
underground experiments, or the ${\tilde W}_1$ is observable at LEP2 (or the
Tevatron).  One may in
fact show further for the entire parameter space $m_0$, $m_{{\tilde g}} <
1$~TeV,
$M_{H_3}/M_G < 10$, that if $\tau (p \rightarrow \overline{\nu} K) > 1.5 \times
10^{33}$
yr (which is testable at Super Kamiokande and ICARUS) then either $m_h <
95$~GeV or
$m_{{\tilde W}_1} < 100$~GeV i.e. either the light Higgs or the light ${\tilde
W}_1$
should be observed at LEP200 \cite{54}.  Thus models of this type can be
experimentally
checked in the near future even before the SSC or LHC are operating.

\vglue 0.5cm
{\elevenbf \noindent 17.  Model Without Proton Decay:  Flipped $SU(5) \times
U(1)$.}
\vglue 0.4cm

As discussed in Sec.~14, proton decay is a normal feature of grand unified
models and one
must do something special in order to suppress it.  One natural way of doing
this arises
in the flipped $SU(5) \times U(1)$ group \cite{32}.  Here particles are
assigned to the 10
and $\overline{5}$ representations of $SU(5)$ but with $u \leftrightarrow
d$~and~$e
\leftrightarrow \nu$ interchanges.  There exists than a $\nu_R$ state, and
$e_R^c$ is put
into an $SU(5)$ singlet representation.  The $p \rightarrow \overline{\nu} K^+$
amplitude
is then suppressed by a factor $O (M_Z/M_G)$ and is hence negligible.

\begin{figure}[htb]
\vspace{3.0in}
\caption{$\alpha_i (Q)$~for~$\alpha_3 (M_Z) = 0.118$ in the flipped $SU(5)
\times U(1)$
model with additional triplet and doublet states.  The dotted curves are where
unification would be without these additional states.}
\label{16}
\end{figure}

One difficulty with this model is that it is not fully unified at the Gut scale
as $G$ is
still a product group.  It has been proposed that unification in this model
should
actually be at the Planck scale (as one would want in string theory) rather
than at the
Gut scale \cite{55}.  This can be achieved by adding an additional pair of 10
and
$\overline{10}$ representations with masses arranged to delay unification until
$10^{18}$~GeV.  For $\alpha_3 (M_Z) = 0.118$ one needs a right $(R)$ handed
color triplet
and left $(L)$ handed doublet at masses $m_R = 4.5 \times 10^6$~GeV and $m_L =
4.1 \times
10^{12}$~GeV.  This is illustrated in Fig.~\ref{16} \cite{55}.  For the strict
No-scale model, where
$A_0 = m_0 = 0 = B_0$, radiative breaking implies a relation between $\tan
\beta$~and~$m_{{\tilde
g}}$.  One finds that solutions exist for $\mu > 0$ only if $m_t~\l~135$~GeV,
and for $\mu < 0$ only
if $m_t~\r~140$~GeV.  This is illustrated in Fig.~\ref{17}.  Also for $\mu > 0$
one has $m_h <
105$~GeV, while for $\mu < 0$ one has $m_h > 100$~GeV.  Since the No-scale
model has
fewer parameters, it can make more precise predictions.  Squark, gluino and
slepton masses obey
the following relations:  $m_{{\tilde q}} \approx m_{{\tilde g}}$, $m_{{\tilde
e}_L}
\approx m_{{\tilde \nu}} \approx 0.3 m_{{\tilde g}}$, $m_{{\tilde e}_R} \approx
0.18
m_{{\tilde g}}$.  Since the sleptons can be quite light, the tri-lepton SUSY
signature
$[p\overline{p} \rightarrow {\tilde W}_1 + {\tilde Z}_2 + X \rightarrow (\ell_1
\overline{\nu}_1
{\tilde Z}_1) + (\ell_2 \overline{\ell}_2 {\tilde Z}_1) + X]$ should be
accessible to the Tevatron
\cite{55}.  It is interesting that these predictions are quite different from
the $SU(5)$ models,
and so the two approaches are experimentally distinguishable.

\begin{figure}[htb]
\vspace{2.25in}
\caption{Strict No-scale model with unification arranged to be at
$10^{18}$~GeV.  Note
that for $\mu > 0$, the $m_{{\tilde g}} - \tan \beta$ relation is double valued
with
$m_t~\l~135$~GeV, while for $\mu < 0$, the model predicts $m_t~\r~140$~GeV.}
\label{17}
\end{figure}

\vglue 0.5cm
{\elevenbf \noindent 18.  Cosmological Constraints}
\vglue 0.4cm

The recent COBE data has given strong support to the Big Bang cosmology and to
the
inflationary scenario.  COBE has measured the cosmic microwave background
roughly at
year 300,000 and finds large scale fluctuations in temperature of size $\Delta
T/T
\simeq 10^{-5}$.  We define the quantity $\Omega = \rho/\rho_c$ where $\rho$ is
the
mass density of the universe and $\rho_c$ is the critical mass density need to
close
the universe, i.e.

\begin{eqnarray}
\rho_c &=& 3 H_0^2/8\pi G_N \nonumber \\
&=& 1.88 h_0^2 \times 10^{-29} gm/cm^3 \label{17.1}
\end{eqnarray}

\noindent
Here $H_0$ is the current Hubble constant, $h_0$ is the Hubble constant in
units of 100
km/sec$Mpc$, and $G_N$ is the Newtonian constant.  The inflationary scenario
then
requires $\Omega = 1$.

Recent measurements of $H_0$ give a range of

\begin{equation}
0.65 < h_0 < 0.75 \label{17.2}
\end{equation}

\noindent
The visable (baryonic) matter in the universe is estimated at
$\Omega_B~\l~0.1$.  Thus
there must be a considerable amount of dark matter present in the universe if
the
inflationary scenario is valid.  A popular model is to assume a mix of cold
dark matter
(CDM) and hot dark matter in the ratio of $\Omega_{CDM}/\Omega_{HDM} \simeq 2$
(i.e.
$\Omega_{CDM} \simeq 0.6$, $\Omega_{HDM} \simeq 0.3$).  This gives a good fit
to the
COBE and other cosmological data.  [We note one possible problem with this
picture
having to do with the age of the universe to:  For $\Omega = 1$, one has $t_0 =
6.5
\times 10^9$ yr/$h_0$ or $t_0 = 10$ billion years for $h_0 = 0.65$.  However,
astrophysical calculations of the age of globular clusters gives a $2\sigma$
spread of
($13~\l$~to~$\l~20$) billion years and one would need $h_0~\l~0.5$ to satisfy
this
bound.]

It is commonly assumed that the HDM are massive $\tau$ neutrinos (left over
from the
Big Bang).  The contribution of massive neutrinos to $\Omega$ is $\Omega_{\nu}
=
(1/4 h_0^2)~(m_{\nu_{\tau}}/23~{\rm eV})$ giving $m_{\nu_{\tau}} \approx
(10)$~eV for
$\Omega_{\nu} = 0.3$.  Two experiments at CERN, CHORUS and NOMAD, plan to
measure
$\nu_{\mu} - \nu_{\tau}$ neutrino oscillations and should be able to detect a
$\nu_{\tau}$ mass in this range (if the mixing angle $\theta$ obeys $\sin^2
2\theta~\r~2 \times 10^{-3}$).

For SUSY models with $R$ parity, the lightest supersymmetric particle (LSP) is
stable.  For most models the LSP is the ${\tilde Z}_1$.  The ${\tilde Z}_1$
particles
created at the time of the Big Bang are hence a natural candidate for CDM.  In
general, in order not to overclose the universe, a minimum requirement on
models with
$R$ parity is

\begin{equation}
\Omega_{{\tilde Z}_1} h_0^2 < 1 \label{17.3}
\end{equation}

\noindent
If the ${\tilde Z}_1$ is the CDM one would need $\Omega_{{\tilde Z}_1} h_0^2
\approx
0.25 - 0.30$.

While the ${\tilde Z}_1$ that were initially created cannot decay, they can
annihilate in pairs in the early universe.  The main mechanisms are via
$h$~and~$Z$~s-channel poles and squark and slepton $t$-channel poles as shown
in
Fig.~\ref{18}.

\begin{figure}[htb]
\vspace{2.75in}
\caption{Annihilation processes for ${\tilde Z}_1$ pairs.  ${\tilde f}
=$~squark or
slepton and $f =$~quark or lepton.}
\label{18}
\end{figure}

The calculation of $\Omega_{{\tilde Z}_1} h_0^2$ follows well known procedures
\cite{56} which we
briefly summarize now.  If $n =$~number/vol of ${\tilde Z}_1$'s, then $n$ obeys
a Boltzman equation
in the early universe:

\begin{equation}
\frac{dn}{dt}~= - 3 H n - <\sigma v> (n^2 - n_0^2) \label{17.4}
\end{equation}

\noindent
where $H$ is the Hubble constant at time $t$, $n_0$ is the value of $n$ at
thermal
equilibrium, $\sigma$ is the annihilation cross section, and $v$ is the
${\tilde Z}_1$
relative velocity.  The first term on the r.h.s. of Eq.~(\ref{17.4}) represents
the
decrease in $n$ due to the expansion of the universe (increase in volume).  The
collision term assumes that the annihilation products go quickly into
equilibrium with
the thermal background so that the inverse scattering process is governed by
$n_0$.
The expression $< >$ means thermal average, and since annihilation generally
occurs at
temperatures when the ${\tilde Z}_1$ are non-relativistic, one may use the
Boltzman
distribution to calculate it:

\begin{equation}
<\sigma v> = \int_0^{\infty} dv v^2 (\sigma v) e^{-v^2/4x}/\int_0^{\infty} dv
v^2
e^{-v^2/4x} \label{17.5}
\end{equation}

\noindent
where $x \equiv kT/m_{{\tilde Z}_1}$~and~$T$ is the temperature.

It is convenient to introduce $f(x) = n/T^3$, and replace $t$~by~$T$ as the
independent variable.  Then

\begin{equation}
\frac{df}{dx}~= \frac{m_{\tilde Z_1}}{k^3}~(\frac{8\pi^3 N_f
G_N}{45})^{-\frac{1}{2}}~<\sigma v>~(f^2 - f_0^2) \label{17.6}
\end{equation}

\noindent
where $N_f$ is the number of degrees of freedom at temperature $T$~and~$f_0 =
n_0/T^3$:

\begin{equation}
f_0 = \frac{2k^3}{(2\pi x)^{3/2}}~e^{-1/x} \label{17.7}
\end{equation}

\noindent
At large $T$, the ${\tilde Z}_1$ is in thermal equilibrium with the background
and so
$f = f_0$.  However, when the annihilation rate becomes smaller than the
expansion
rate, the ${\tilde Z}_1$ decouples from the background.  This occurs at the
``freeze
out'' temperature $T_f$.  Then $f$ obeys

\begin{equation}
\frac{df}{dx}~= \frac{m_{{\tilde Z}_1}}{k^3}~(\frac{8\pi^3 N_f
G_N}{45})^{-\frac{1}{2}}~<\sigma v> f^2;~~x \leq x_f \equiv k T_f/m_{{\tilde
Z}_1}
\label{17.8a}
\end{equation}

\noindent
with boundary condition

\begin{equation}
f (x_f) = f_0 (x_f) \label{17.8b}
\end{equation}

\noindent
Eqs.~(\ref{17.8a}),(\ref{17.8b}) allows one to solve for $x_f$:

\begin{equation}
x_f^{-1} \cong ln \bigg[x_f^{1/2} <\sigma v>_{x_f} m_{{\tilde Z}_1}
\sqrt{\frac{45}{N_f G_N}}\bigg] \label{17.9}
\end{equation}

\noindent
In general $x_f$ is small, i.e. $x_f \approx 1/20$ (showing that the freeze out
does
indeed occur in the non-relativistic domain) and Eq.~(\ref{17.9}) can be solved
for
$x_f$ by iteration.

Integrating Eq.~(\ref{17.8a}) from $x_f$~to~$x_0 = kT_0/m_{{\tilde Z}_1} \cong
0$,
where $T_0$ is the present temperature, one obtains the relic density at
present time
$t_0$:

\begin{equation}
\rho_{{\tilde Z}_1} = 4.75 \times 10^{-40} (\frac{T_{{\tilde
Z}_1}}{T_{\gamma}})^3~(\frac{T_{\gamma}}{2.75})^3~\frac{(N_f)^{1/2}}{J
(x_f)}~g/cm^3
\label{17.20}
\end{equation}

\noindent
Here $T_{\gamma}$ is the current microwave background temperature, and $J$ is

\begin{equation}
J (x_f) = \int_0^{x_f} dx <\sigma v> (x) {\rm GeV}^{-2} \label{17.21}
\end{equation}

\noindent
{}From Eq.~(\ref{17.1}) we have then

\begin{equation}
\Omega_{{\tilde Z}_1} h_0^2 = 2.53 \times 10^{-11} (\frac{T_{{\tilde
Z}_1}}{T_{\gamma}})^3~(\frac{T_{\gamma}}{2.75})^3~\frac{(N_f)^{1/2}}{J (x_f)}
\label{17.22}
\end{equation}

\noindent
In Eqs.~(\ref{17.20}) and (\ref{17.22}), the factor $(T_{{\tilde
Z}_1}/T_{\gamma})^3$
arises from the reheating of the photon temperature due to the annihilation of
particles of mass less than $k T_f$ \cite{56}.  For a ${\tilde Z}_1$ with mass
of
(20-50) GeV, one has \cite{57}

\begin{equation}
(\frac{T_{\gamma}}{T_{{\tilde Z}_1}})^3~\cong 18.5;~~N_f \cong 289.5
\label{17.23}
\end{equation}

\noindent
Eqs. (\ref{17.22}) and (\ref{17.23}) allows an explicit calculation of
$\Omega_{{\tilde Z}_1} h_0^2$ once the double integral $J (x_f)$ is computed.

We come now to an important fine point.  Since $x_f$ is small and hence we are
in the
non-relativistic regime, it had generally been thought previously that one
could
expand $\sigma v$ in a power series in $v^2$, $\sigma v = a + b v^2/6 +
\cdots$.  Then
the thermal average becomes trivial to perform i.e. $<\sigma v> \cong a + bx$
(since
$<v^2> = 6kT/m_{{\tilde Z}_1}$).  However, as has been pointed out \cite{58}
this
approximation can breakdown badly near a pole or threshold.  The breakdown is
particularly serious for the physical case at hand due to the narrowness of the
$h$~and~$Z$ poles \cite{59}, invalidating a good deal of the earlier analysis.
To see
what the problem is, consider the $h$-pole where $\sigma v$ has the
Breit-Wigner form

\begin{equation}
\sigma v \cong A_h \frac{v^2}{[(v^2 - \varepsilon_R)^2 + \gamma_R]}
\label{17.24}
\end{equation}

\noindent
Here $\varepsilon_R = (m_h^2 - 4m_{{\tilde Z}_1}^2)/m_{{\tilde Z}_1}^2$,
$\gamma_R =
m_h \Gamma_h/m_{{\tilde Z}_1}^2$~and~$A_h$ is a constant \cite{60}.  The $h$
width is
$\Gamma_h \simeq 2.5 \times 10^{-3}$~GeV and hence $\gamma_R$ is very small.
When one
thermally averages, one ``smears'' $v^2$, and if $\varepsilon_R > 0$ (i.e. $2
m_{{\tilde Z}_1} < m_h$) then when $v^2 \approx \varepsilon_R$ the integrand
becomes
very large and one can get a large enhancement in $<\sigma v>$.  This will
modify the
value of $x_f$ and produce  changes in $\Omega_{{\tilde Z}_1} h_0^2$ by factors
as
large as 1,000.

In general one may calculate the integral for $<\sigma v>$ numerically without
much
trouble and thus obtain the correct freeze out parameter $x_f$.  It is more
complicated to calculate the double integral of $J (x_f)$ due to the singular
nature
of the pole.  A convenient procedure is to do the $x$ integral anatylically
first to
obtain

\begin{eqnarray}
J_h &=& \frac{A_h}{m_{{\tilde Z}_1}^4 2\pi^{1/2}}~\int_0^{\infty} d \xi
e^{-\xi}
\xi^{-\frac{1}{2}} \bigg\{\frac{1}{2}~ln \bigg[\frac{(4\xi x_f -
\varepsilon_R)^2 +
\gamma_R^2}{\varepsilon_R^2 + \gamma_R^2}~\bigg] \nonumber \\
&+& \frac{\varepsilon_R}{\gamma_R}~\bigg[\tan^{-1} (\frac{4\xi x_f -
\varepsilon_R}{\gamma_R})~+ \tan^{-1}
(\frac{\varepsilon_R}{\gamma_R})~\bigg]\bigg\}
\label{17.25}
\end{eqnarray}

\noindent
The remaining integral can then be done numerically without difficulty.  A
similar
analysis can be carried out for the $Z$ pole contributions.  For the
$t$-channel
exchanges of sfermions, the expansion $\sigma v \cong a + b v^2/6$ is good
approximation as one is not near a singularity.

\begin{figure}[htb]
\vspace{2.5in}
\caption{$\Omega_{approx}/\Omega$ for $m_0 = 700$~GeV, $\tan \beta = 2.25$,
$\mu > 0$
for $m_t = 110$~GeV, $A_t/m_0 = -0.8$ (dashed), $m_t = 125$~GeV, $A_t/m_0 = -
0.4$
(solid), and $m_t = 140$~GeV, $A_t/m_0 = 0.0$ (dotted).  The $h$~and~$Z$ poles
occur
where the curves steeply go from positive to negative.}
\label{19}
\end{figure}

Fig.~\ref{19} shows the ratio of the approximate value of $\Omega$ to the
correct value as a
function of $m_{{\tilde g}}$.  One sees that the approximate value of $\Omega$
makes a large error in
the region prior to each pole.  From Fig.~\ref{20} this is precisely the region
where the correct
$\Omega_{{\tilde Z}_1} h_0^2 < 1$ and hence cosmologically acceptable.  One can
also
see from Fig.~22, that the correct cosmologically allowed region is both larger
and in
a different $m_{{\tilde g}}$ region then for the approximate calculation,
showing that
the approximate analysis can lead to significant error.

\begin{figure}[htb]
\vspace{2.5in}
\caption{$\Omega h_0^2$ (solid) and $(\Omega h_0^2)_{approx}$ (dashed) for $m_t
=
125$~GeV, $m_0 = 700$~GeV, $\tan \beta = 1.88$, $A_t/m_0 = 0.5$, $\mu > 0$ in
the
vacinity of the $h$ pole.}
\label{20}
\end{figure}

We now turn to consider the effect on the parameter space implied by the
cosmological
constraint of Eq. (\ref{17.3}).  For the flipped No-scale model $(m_{1/2} \gg
m_0)$
one generally finds $\Omega h_0^2 < 1$, so no additional condition is implied
\cite{55}.  For models where $m_0$ is not zero, one generally finds $\Omega
h_0^2$ to be
large, e.g. $O (10)$ except, as illustrated in Figs.~21 and 22 in the region
prior to
the $h$~and~$Z$ poles.  Thus for the $h$ pole, this would imply an allowed
region
where $2 m_{{\tilde Z}_1} < m_h$ and since the ${\tilde Z}_1$ and gluino masses
scale,
to a band of allowed gluino masses.  The cosmological constraint then reduces
the five dimensional
parameter space $(m_{{\tilde g}}$, $m_0$, $A_0$, $\tan \beta$~and~$m_t)$ to a
five dimensional shell
with thickness in $m_{{\tilde g}}$ of $\approx (5 - 125)$~GeV thick.  This is
illustrated in Figs.~23 and 24.  In Fig.~\ref{21}, the allowed region between
the two solid
lines comes from the region where $2 m_{{\tilde Z}_1} < m_h$ while the region
between the dashed
curves comes from the $Z$ pole contribution.  Sometimes, these two regions
overlap, leading to one
large allowed region.  An example of this is illustrated in Fig.~\ref{22}.

\begin{figure}[htb]
\vspace{2.5in}
\caption{Allowed $m_{{\tilde g}}$ region as a function of $A_t$ for $m_t =
125$~GeV,
$m_0 = 600$~GeV, $\tan \beta = 1.73$, $\mu > 0$.  The allowed region is that
between
the two solid lines (from the $h$ pole) and that between the two dashed lines
(from
the $Z$ pole).}
\label{21}
\end{figure}

\begin{figure}[htb]
\vspace{2.5in}
\caption{Allowed $m_{{\tilde g}}$ region as a function of $A_t$ for $m_t =
140$~GeV,
$m_0 = 700$, $\tan \beta = 2.75$, $\mu > 0$.}
\label{22}
\end{figure}

As a final point we consider the combined relic density constraint
Eq.~(\ref{17.3})
and proton decay constraint.  Both of these could operate in a supergravity Gut
model
with $R$ parity and $SU(5)$-type proton decay.  The regions allowed by relic
density
constraint in Figs.~23 and 24 also obey the current proton decay experimental
bounds
with $M_{H_3}/M_G < 6$~$(B < 600~{\rm GeV}^{-1})$.  Fig.~\ref{23} shows the
value of the proton decay
amplitude $B$ for the region allowed by the relic density constraint for $m_0 =
700$~GeV as a
function of $A_t$ (as one varies all other parameters) for $m_t = 110$~GeV, 125
GeV and 140 GeV.
Note that the allowed region increases with $m_t$.  The region below the
horizontal line is that
allowed by proton decay for $M_{H_3}/M_G < 6$.  For $M_{H_3}/M_G < 10$ this
line would be at $B =
1000$~GeV$^{-1}$.  The simultaneous relic density and proton decay constraints
reduce
the allowed parameter space.  However, a considerable region is still allowed.
For
$M_{H_3}/M_G < 6$, one finds that the combined constraints imply $m_h <
105$~GeV,
$m_{{\tilde W}_1} < 100$~GeV, $m_{{\tilde Z}_1}~\l~50$~GeV and
$m_t~\l~165$~GeV.
These bounds are relaxed somewhat if one raises $M_{H_3}$~to~$M_{H_3}/M_G < 10$
(and
there are small corners in parameter space where $m_{{\tilde W}_1} \approx
300$~GeV
and $m_{{\tilde Z}_1} \approx 150$~GeV).

\begin{figure}[htb]
\vspace{2.75in}
\caption{Proton decay amplitude $B$ as a function of $A_t$ for $m_0 = 700$~GeV.
The
dashed line is for $m_t = 110$~GeV, the solid line for $m_t = 125$~GeV and the
dotted
line for $m_t = 140$~GeV.  (The gap in the central region for $m_t = 110$~GeV
is due
to the requirement $m_h > 60$~GeV.)  The region allowed by the relic density
constraint is between the upper and lower curves.}
\label{23}
\end{figure}

\vglue 0.5cm
{\elevenbf \noindent 19.  Concluding Remarks}
\vglue 0.4cm

Several proposals currently exist to remedy the theoretical weaknesses of the
Standard
Model.  Supersymmetry offers a solution to the gauge hierarchy problem in that
it offers
a natural way for the quadratic divergences of the Higgs self energy to cancel.
However, it does this at the expense of greatly enlarging the theoretical
structure.
Thus the number of particles in the low energy domain is doubled.  Further, in
order to
obtain a phenomenologically acceptable way of spontaneously breaking
supersymmetry, one
is lead to promoting it to a local gauge symmetry i.e. supergravity.  Here,
supersymmetry breaking arises from Planck scale interactions, implying that one
is
dealing with a theory whose natural energy scale is high above the electroweak
scale.
The recent data, suggesting the validity of supersymmetric grand unification
supports
this view and implies also the existance of a desert between the two scales.
Thus the
simplest way of consistently implementing the supersymmetric solution to the
gauge
hierarchy problem is within the framework of supergravity grand unification.

Such an enlargement of the theoretical structure would be unreasonable were it
not for
two things.  First, one can control the unknown nature of the Planck scale
interactions
which give rise to supersymmetry breaking by imposing simple phenomenological
constraints i.e. the absence of flavor changing neutral interactions, and the
existance
of grand unification.  This then allows one to parameterize the supersymmetry
breaking
sector in terms of only four unknown constants.  Hence the masses, production
cross
sections, decay widths etc. of the 32 new SUSY particles can be obtained in
terms of
these four new parameters.  Second, throughout the entire energy domain, the
theory
stays within the perturbative domain.  As a consequence, explicit calculations
of the
predictions of the theory can be made and the theoretical extropolations made
can be
experimentally tested.  How to calculate some of the consequences of
supergravity grand
unified models is given in this report.  Some of the more natural models should
be
tested experimentally in the relatively near future, with detailed tests
requiring the
SSC or LHC.

There are a number of important questions not explained by supergravity models.
Perhaps
the most important is the nature of the quark and lepton mass matrices, and
this may
require Planck scale physics to understand.  Superstring theory has the
capability of
deducing Yukawa couplings, though the huge number of candidate string vacua and
the
problem of supersymmetry breaking in string theory has inhibited progress in
this
direction.  There have been, however, recent phenomenological analyses of mass
matrices
\cite{61} based on the idea that the generational hierarchy is scaled by
$M_G/M_{P\ell}$,
implying a Planck physics origin of at least the first two generations mass.
Finally we
mention the important question of the origin of CP violation.  Supergravity
models
appear to shed no light on this question either.  CP violation may also be a
Planck
physics phenomena (in the Standard Model CP violation is characterized by a
phase in the
CKM matrix and hence is related to the quark mass matrix).  More experimental
data is
obviously needed to understand this phenomena better.

\vglue 0.5cm
{\elevenbf \noindent  Acknowledgements}
\vglue 0.4cm

This work was supported in part by National Science Foundation Grants Nos.
PHY-916593 and
PHY-9306906.  One of us (R.A.) would like to thank the Superconducting Super
Collider Laboratory
for its kind hospitality during the writing of this report and where part of
the work described
here was done.

\vglue 0.5cm
{\elevenbf \noindent References and Footnotes}
\vglue 0.4cm


\begin{thebibliography}{61}
\bibitem{1} G. Altarelli, Les Rencontres de Physique de la Valee D'Aoste, La
Thuile, 1993.

\bibitem{2} P. Langacker, Lectures at TASI-92, Boulder, 1992, UPR-O55T, March,
1993.  The
prediction for $m_t$ is $m_t = 150^{+ 19 + 15}_{- 24 - 20}$~GeV where the
central value is for the
Higgs mass $m_H = 300$~GeV and the second error is from assuming 60 GeV $< m_H
< 1000$~GeV (smaller
$m_H$ giving smaller $m_t$).  The best fit occurs for $m_H = 60$~GeV and the
upper limit on $m_t$
is $m_t < 208$~GeV at 99\% CL (which occurs at $m_H = 1000$~GeV).

\bibitem{3} P. Langacker, Proc. PASCOS 90-Symposium, Eds. P. Nath and S.
Reucroft (World
Scientific, Singapore 1990); J. Ellis, S. Kelley and D. V. Nanopoulos, Phys.
Lett. {\bf 249B}, 441
(1990); {\bf B260}, 131 (1991); U. Amaldi, W. de Boer and H. Furstenau, Phys.
Lett. {\bf 260B}, 447
(1991); F. Anselmo, L. Cifarelli, A. Peterman and A. Zichichi, Nuov. Cim. {\bf
104A}, 1817 (1991);
{\bf 115A}, 581 (1992).

\bibitem{4} Particle Data Group, Phys. Rev. {\bf D45}, Part 2 (June, 1992).

\bibitem{5} For a discussion of global supersymmetry see e.g. J. Wess and J.
Bagger, Supersymmetry
and Supergravity (Princeton University Press, Princeton 1983).

\bibitem{6} Our notation is as follows:  ${\rm diag}~\eta_{\mu\nu} = (-1, 1, 1,
1)$, $P^0 = H
=$~Hamiltonian $(P_0 = - P^0)$, $P^i =$~momentum, and $\gamma^{\mu}$ obey
$\{\gamma^{\mu},
\gamma^{\nu}\} = - 2 \eta^{\mu\nu}$.  In the Majorana representation
$\gamma^{\mu \ast} = -
\gamma^{\mu}$, ${\tilde \gamma}^0 = - \gamma^0$, ${\tilde \gamma}^i = +
\gamma^i$~and~$\gamma^{5\ast} = - \gamma^5$, ${\tilde \gamma}^5 = - \gamma^5$.

\bibitem{7} R. Haag, J. Lopuszanski and M. Sohnius, Nucl. Phys. {\bf B88}, 257
(1975).

\bibitem{8} Thus one can consider extended supersymmetries based on
$Q_{\alpha}^A$,$A = 1 \cdots N$
where $\{Q_{\alpha}^A, Q_{\beta}^{B \dagger}\} = - 2 (P_L \gamma^{\mu}
\gamma^0)_{\alpha\beta}
\delta^{AB} P_{\mu}$ and add central charges.  Algebras with central charges,
however, do not allow
massless states.

\bibitem{9} S. Weinberg, Phys. Lett. {\bf B62}, 111 (1976); E. Witten, Nucl.
Phys. {\bf B202}, 253
(1982).

\bibitem{10} M. Grisaru, M. Ro\^cek and W. Seigel, Nucl. Phys. {\bf B159}, 429
(1979).

\bibitem{11} Note that by convention, all chiral multiplets are left-handed
e.g. $(z, \chi_L)$.
Thus instead of using right handed quark or leptons one must use the
left-handed conjugate spinors
as basic variables e.g. $(u^C)_L = (u_R)^C$.  The corresponding multiplet is
denoted by $({\tilde
u}^C, u^C_L)$.  (In the Majorana representation one has $\chi^C =
\chi^{\dagger}$.)

\bibitem{12} L. Giradello and M. T. Grisaru, Nucl. Phys. {\bf B194} 65 (1982).

\bibitem{13} In the effective potential, all fields are spin zero bosons, and
we will from now on
drop the tilde notation for squarks and sleptons.

\bibitem{14} F. Abbe et al., Phys. Rev. Lett. {\bf 69}, 3439 (1992).

\bibitem{15} M. Davier, Proc. Lepton-Photon High Energy Physics Conference,
Geneva, 1991, Eds. S.
Hegarty, K. Potter, E. Quercigh (World Scientific, Singapore (1991).

\bibitem{16} S. Franchiotti, B. Kniehl and A. Sirlin, Phys. Rev. {\bf D48}, 307
(1993).

\bibitem{17} P. Langacker, Ref. [2].

\bibitem{18} H. Bethke, XXVI Proc. Conference on High Energy Physics, Dallas
1992, Ed. J. Sanford,
AIP Conf. Proc. No. 272 (1993) G. Altarelli, talk at Europhysics Conference, on
High Energy
Physics, Marseille, August 1993.

\bibitem{19} M. B. Einhorn and D. R. T. Jones, Nucl. Phys. {\bf B196}, 475
(1982).

\bibitem{20} G. Altarelli, Ref. [18].

\bibitem{21} D. Freedman, S. Ferrara and P. van Nieuwenhuizen, Phys. Rev. {\bf
D13}, 3214 (1976);
S. Deser and B. Zumino, Phys. Lett. {\bf B62}, 335 (1976).

\bibitem{22} A. H. Chamseddine, R. Arnowitt and P. Nath, Phys. Rev. Lett. {\bf
29}, 970 (1982); E.
Cremmer, S. Ferrara, L. Girardello and A. van Proeyen, Phys. Lett. {\bf 116B},
231 (1982); Nucl.
Phys. {\bf B212}, 413 (1983); E. Witten and J. Bagger, Nucl. Phys. {\bf B222},
125 (1983).

\bibitem{23} For a review see P. Nath, R. Arnowitt and A. H. Chamseddine,
``Applied \protect$N = 1\protect$
Supergravity'' (World Scientific, Singapore (1984); H. P. Nilles, Phys. Rep.
{\bf 110}, 1 (1984).

\bibitem{24} More precisely, the general form of the Kahler potential is 
\protect$d = z_i z_i^{\dagger} + f(z_i) + f^{\dagger} (z_i) + \cdots\protect$.  
One may then remove the function \protect$f (z_i)\protect$ [by a Kahler
transformation Eq.~({\ref{8.2}})] into the superpotential to obtain the form
given in text.

\bibitem{25} J. Polonyi, Univ. of Budapest Rep. No. KFKI-1977-93 (1977).

\bibitem{26} H. P. Nilles, Phys. Lett. {\bf B115}, 193 (1981); S. Ferrara, L.
Girardello and H. P.
Nilles, Phys. Lett. {\bf B125}, 457 (1983).

\bibitem{27} A. H. Chamseddine, R. Arnowitt and P. Nath, Phys. Rev. Lett. {\bf
49}, 970 (1982); L.
E. Iba\~nez, Phys. Lett. {\bf B118}, 73 (1982); J. Ellis, D. V. Nanopoulos and
K. Tamvakis, Phys.
Lett. {\bf B121}, 123 (1983).

\bibitem{28} A full derivation of this result does not exist in the literature.
See however, A.
H. Chamseddine, R. Arnowitt and P. Nath, Ref. [27]; R. Barbieri, S. Ferrara and
C. A. Savoy, Phys.
Lett. {\bf B119}, 343 (1982); L. Hall, J. Lykken and S. Weinberg, Phys. Rev.
{\bf D27}, 2359
(1983); P. Nath, R. Arnowitt and A. H. Chamseddine, Nucl. Phys. {\bf B227}, 121
(1983); S. Soni
and A. Weldon, Phys. Lett. {\bf B126}, 215 (1983).

\bibitem{29} E. Witten, Nucl. Phys. {\bf B177}, 477 (1981); {\bf B185}, 513
(1981); S. Dimopoulos
and H. Georgi, Nucl. Phys. {\bf B193}, 150 (1981); N. Sakai, Zeit. f. Phys.
{\bf C11}, 153 (1981).

\bibitem{30} B. Grinstein, Nucl. Phys. {\bf B206}, 387 (1982); H. Georgi, Phys.
Lett. {\bf B115},
380 (1982).

\bibitem{31} K. Inoue, A. Kakuto and T. Tankano, Prog. Theor. Phys. {\bf 75},
664 (1986); A.
Anselm and A. Johansen, Phys. Lett. {\bf B200}, 331 (1988); A. Anselm, Sov.
Phys. JETP {\bf 67},
663 (1988); R. Barbieri, G. Dvali and A. Strumia, Nucl. Phys. {\bf B391}, 487
(1993).

\bibitem{32} I. Antoniadis, J. Ellis, J. Hagelin and D. V. Nanopoulos, Phys.
Lett. {\bf B208}, 209
(1988).

\bibitem{33} K. Inoue et al., Prog. Theor. Phys. {\bf 68}, 927 (1982); L.
Iba\~nez and G. G. Ross,
Phys. Lett. {\bf B110}, 227 (1982); L. Alvarez-Gaum\'e, J. Polchinski and M. B.
Wise, Nucl. Phys.
{\bf B250}, 495 (1983); J. Ellis, J. Hagelin, D. V. Nanopoulos and K. Tamvakis,
Phys. Lett. {\bf
B125}, 2275 (1983); L. E. Iba\~nez and C. Lopez, Phys. Lett. {\bf B128}, 54
(1983); Nucl. Phys.
{\bf B233}, 545 (1984), L. E. Iba\~nez, C. Lopez and C. Mu\~nos, Nucl. Phys.
{\bf B256}, 218
(1985); J. Ellis and F. Zwirner, Nucl. Phys. {\bf B338}, 317 (1990).

\bibitem{34} A. H. Chamseddine, R. Arnowitt and P. Nath Ref. [27].

\bibitem{35} S. Coleman and E. Weinberg, Phys. Rev. {\bf D7}, 1888 (1973); S.
Weinberg, {\bf D7},
2887 (1973).

\bibitem{36} G. Gamberini, G. Ridolfi and F. Zwirner, Nucl. Phys. {\bf B331},
331 (1990).

\bibitem{37} R. Arnowitt and P. Nath, Phys. Rev. {\bf D46}, 3981 (1992).

\bibitem{38} J. M. Frere, D. Jones and S. Raby, Nucl. Phys. {\bf B222}, 1
(1983); M. Claudson, L.
Hall and I. Hinchliffe, Nucl. Phys. {\bf B288}, 501 (1983); M. Drees, M. Gluck
and K. Brassie,
Phys. Lett. {\bf B157}, 164 (1985).

\bibitem{39} Y. Ikada, M. Yamaguchi and T. Yanagida, Prog. Theor. Phys. {\bf
85}, 1 (1991); J.
Ellis, G. Ridolfi and E. Zwirner, Phys. Lett. {\bf B257}, 83 (1991); H. E.
Haber and R. Hampling,
Phys. Rev. Lett. {\bf 66}, 1815 (1991).

\bibitem{40} J. Ellis, S. Kelley and D. V. Nanopoulos, Nucl. Phys. {\bf B373},
55 (1992); Phys.
Lett. {\bf B287}, 95 (1992); F. Anselmo et al. [3].

\bibitem{41} R. Barbieri and L. J. Hall, Phys. Rev. Lett. {\bf 68}, 752 (1992);
A. Faraggi, B.
Grinstein and S. Meshkov, Phys. Rev. {\bf D47}, 5018 (1993); L. Hall and U.
Sarid, Phys. Rev.
Lett. {\bf 70}, 2673 (1993).

\bibitem{42} P. Langacker and N. Polonsky, Phys. Rev. {\bf D47}, 4028 (1993).

\bibitem{43} D. Ring and S. Urano, unpublished (1992).

\bibitem{44} Particle Data Group, Phys. Rev. {\bf D45}, Part 2 (June 1992).

\bibitem{45} S. Weinberg, Phys. Rev. {\bf D26}, 287 (1982); N. Sakai and T.
Yanagida, Nucl. Phys.
{\bf B197}, 533 (1982); S. Dimopoulos, S. Raby and F. Wilczek, Phys. Lett. {\bf
B112}, 133 (1982);
J. Ellis, D. V. Nanopoulos and S. Rudaz, Nucl. Phys. {\bf B202}, 43 (1982); S.
Chadha and M.
Daniels, Nucl. Phys. {\bf B229}, 105 (1983); B. A. Campbell, J. Ellis and D. V.
Nanopoulos, Phys.
Lett. {\bf B141}, 224 (1984).

\bibitem{46} R. Arnowitt, A. H. Chamseddine and P. Nath, Phys. Lett. {\bf
B156}, 215 (1985); P.
Nath, R. Arnowitt and A. H. Chamseddine, Phys. Rev. {\bf D32}, 2348 (1985).

\bibitem{47} M. B. Gavela et al., Nucl. Phys. {\bf B312}, 269 (1989).

\bibitem{48} R. Arnowitt and P. Nath, Phys. Rev. Lett. {\bf 69}, 725 (1992).

\bibitem{49} P. Nath and R. Arnowitt, Phys. Lett. {\bf B289} 368 (1992).

\bibitem{50} J. F. Gunion and H. E. Haber, Phys. Rev. {\bf D37}, 2515 (1988).

\bibitem{51} P. Nath and R. Arnowitt, Phys. Lett. {\bf B287}, 3282 (1992); K.
Inoue, M. Kawasaki,
M. Yamaguchi and T. Yanagida, Phys. Rev. {\bf D45}, 328 (1992).

\bibitem{52} Y. Totsuka, Proc. XXIV Conf. on High Energy Physics, Munich, 1988,
Eds. R. Kotthaus
and J. H. Kuhn (Springer Verlag, Berlin, Heidelberg, 1989).

\bibitem{53} ICARUS Detector Group, Int. Symposium on Neutrino Astrophysics,
Takayama, 1992.

\bibitem{54} R. Arnowitt and P. Nath,
CTP-TAMU-32/93-NUB-TH-3066/93-SSCL-Preprint-440 (1993).

\bibitem{55} J. L. Lopez, D. V. Nanopoulos and A. Zichichi, CERN-TH 6667
(1993).

\bibitem{56} B. W. Lee and S. Weinberg, Phys. Rev. Lett. {\bf 39}, 165 (1977);
D. A. Dicus, E.
Kolb and V. Teplitz, Phys. Rev. Lett. {\bf 39}, 168 (1977); H. Goldberg, Phys.
Rev. Lett. {\bf
50}, 1419 (1983); J. Ellis, J. S. Hagelin, D. V. Nanopoulos, K. Olive and M.
Srednicki, Nucl.
Phys. {\bf B238}, 453 (1984).

\bibitem{57} K. Olive, D. Schramm, and G. Steigmann, Nucl. Phys. {\bf B180},
497 (1981).

\bibitem{58} K. Griest and D. Seckel, Phys. Rev. {\bf D43}, 3191 (1991); P.
Gondolo and G.
Gelmini, Nucl. Phys. {\bf B360}, 145 (1991).

\bibitem{59} R.  Arnowitt and P. Nath, Phys. Lett. {\bf B299}, 58 (1993) and
Erratum ibid {\bf
B303}, 403 (1993); P. Nath and R. Arnowitt, Phys. Rev. Lett. {\bf 70}, 3696
(1993).

\bibitem{60} J. Lopez, D. V. Nanopoulos and K. Yuan, Nucl. Phys. {\bf B370},
445 (1992); M. Drees
and M. M. Nojiri, Phys. Rev. {\bf D47}, 376 (1993).

\bibitem{61} G. Anderson, S. Dimopoulos, L. Hall and S. Raby, Phys. Rev. {\bf
D47}, 3702 (1993).
\end{thebibliography}
\end{document}